\documentclass[structabstract]{aa}  
\usepackage{graphicx}
\usepackage{txfonts}
\usepackage{natbib}
\usepackage{longtable}
\def\pdeg{\ifmmode $\setbox0=\hbox{$^{\circ}$}\rlap{\hskip.11\wd0 .}$^{\circ}
  \else \setbox0=\hbox{$^{\circ}$}\rlap{\hskip.11\wd0 .}$^{\circ}$\fi}

\begin{document}

\title{The Cygnus Allscale Survey of Chemistry and Dynamical Environments:  CASCADE.}

\subtitle{Overview and first results toward DR20 from the Max Planck IRAM
  Observatory program (MIOP)\thanks{The data
    are available in electronic form at the CDS via anonymous ftp to
    cdsarc.u-strasbg.fr (130.79.128.5) or via
    http://cdsweb.u-strasbg.fr/cgi-bin/qcat?J/A+A/}.}

   \author{H.~Beuther
          \inst{1}
          \and
          F.~Wyrowski 
          \inst{2}
          \and
          K.~M.~Menten
          \inst{2}
          \and
          J.~M.~Winters
          \inst{3}
          \and
          S.~Suri
          \inst{1,4}
          \and
          W.-J.~Kim
          \inst{5}
          \and
          L.~Bouscasse
          \inst{3}
          \and
          C.~Gieser
          \inst{1}
          \and
          M.~Sawczuck
          \inst{1}
          \and
          I.~B.~Christensen
          \inst{2}
          \and
          I.~M.~Skretas
          \inst{2}
}
   \institute{$^1$ Max Planck Institute for Astronomy, K\"onigstuhl 17,
     69117 Heidelberg, Germany, \email{name@mpia.de}\\
     $^2$ Max-Planck-Institut f\"ur Radioastronomie, Auf dem H\"ugel 69, 53121 Bonn, Germany \\
     $^3$ IRAM, 300 rue de la Piscine, Domaine Universitaire de Grenoble, 38406 St.-Martin-d’Hères, France\\
     $^4$ Department of Astrophysics, University of Vienna, T\"urkenschanzstrasse 17,1180 Vienna, Austria\\
     $^5$ I.~Physikalisches Institut, Universit\"at zu K\"oln, Z\"ulpicher Str. 77, 50937 K\"oln, Germany\\
   }

   \date{Version of \today}

\abstract
    {While star formation on large molecular cloud scales and on 
      small core and disk scales has been investigated intensely over
      the past decades, the connection of the large-scale interstellar material with the densest
      small-scale cores has been a largely neglected field.}
    {We wish to understand how the gas is fed from clouds down to
      cores. This covers dynamical accretion flows as well as the
      physical and chemical gas properties over a broad range of
      spatial scales.}
    {Using the IRAM facilities NOEMA and the IRAM 30\,m telescope, we
      mapped large areas (640\,arcmin$^2$) of the archetypical star formation complex
      Cygnus X at 3.6\,mm wavelengths in  line and continuum
      emission. The data were combined and imaged together to cover all accessible spatial
      scales.}
{The scope and outline of The Cygnus Allscale Survey of Chemistry and Dynamical Environments (CASCADE) as part of the Max Planck IRAM Observatory
  Program (MIOP) is presented. We then focus on the first observed
  subregion in Cygnus X, namely the DR20 star formation site, which comprises sources in a 
  range of evolutionary stages from cold pristine gas clumps to more evolved ultracompact H{\sc ii} regions. The data covering cloud to cores
  scales at a linear spatial resolution of $<5000$\,au reveal several
  kinematic cloud components that are likely part of several
  large-scale flows onto the central cores. The temperature structure
  of the region is investigated by means of the HCN/HNC intensity ratio and
  compared to dust-derived temperatures. We find that the deuterated
  DCO$^+$ emission is almost exclusively located toward regions at low
  temperatures below 20\,K. Investigating the slopes of spatial power
  spectra of dense gas tracer intensity distributions (HCO$^+$, H$^{13}
  $CO$^+$, and N$_2$H$^+$), we find
  comparatively flat slopes between $-2.9$ and $-2.6$, consistent with
  high Mach numbers and/or active star formation in DR20.}
{This MIOP large program on star formation in Cyg X provides unique
  new data connecting cloud with core scales. The analysis of the DR20 data presented here highlights the potential of this program to investigate in detail the different physical and chemical aspects and their interrelations from the scale of the natal molecular cloud down to the scale of accretion onto the individual protostellar cores.}


\titlerunning{}

\maketitle
 
\section{Introduction}
\label{intro}

Cloud and star formation are hierarchical processes in which
gas is transported from cloud scales of tens to hundreds of parsec down to
sub-au scales of the inner accretion region. While the gas and dust distributions on large scales can be
studied efficiently via (sub)millimeter and radio wavelength Galactic plane surveys such as the APEX Telescope Large Area Survey of the Galaxy (ATLASGAL; \citealt{schuller2009}) or the Herschel infrared Galactic Plane Survey (HIGAL; \citealt{molinari2016}), the Galactic
Ring survey (GRS; \citealt{jackson2006}), the HI/OH/Recombination line
survey of the Milky Way (THOR; \citealt{beuther2016,wang2020a}), the Global view on Star formation in the Milky Way (GLOSTAR; \citealt{brunthaler2021}) and many others, the small-scale structures
are more typically addressed by millimeter (mm) interferometric studies of
individual target regions. However, the connection of the content of whole clouds
with the smallest-scale structures has been a neglected field
of research. The  main reason is that single-dish instruments do not
resolve the relevant core scales, and mm interferometers did not
allow efficient mapping of large areas on the sky so far. On their own, they cannot image extended structures.

\begin{figure*}[ht] 
\includegraphics[width=0.99\textwidth]{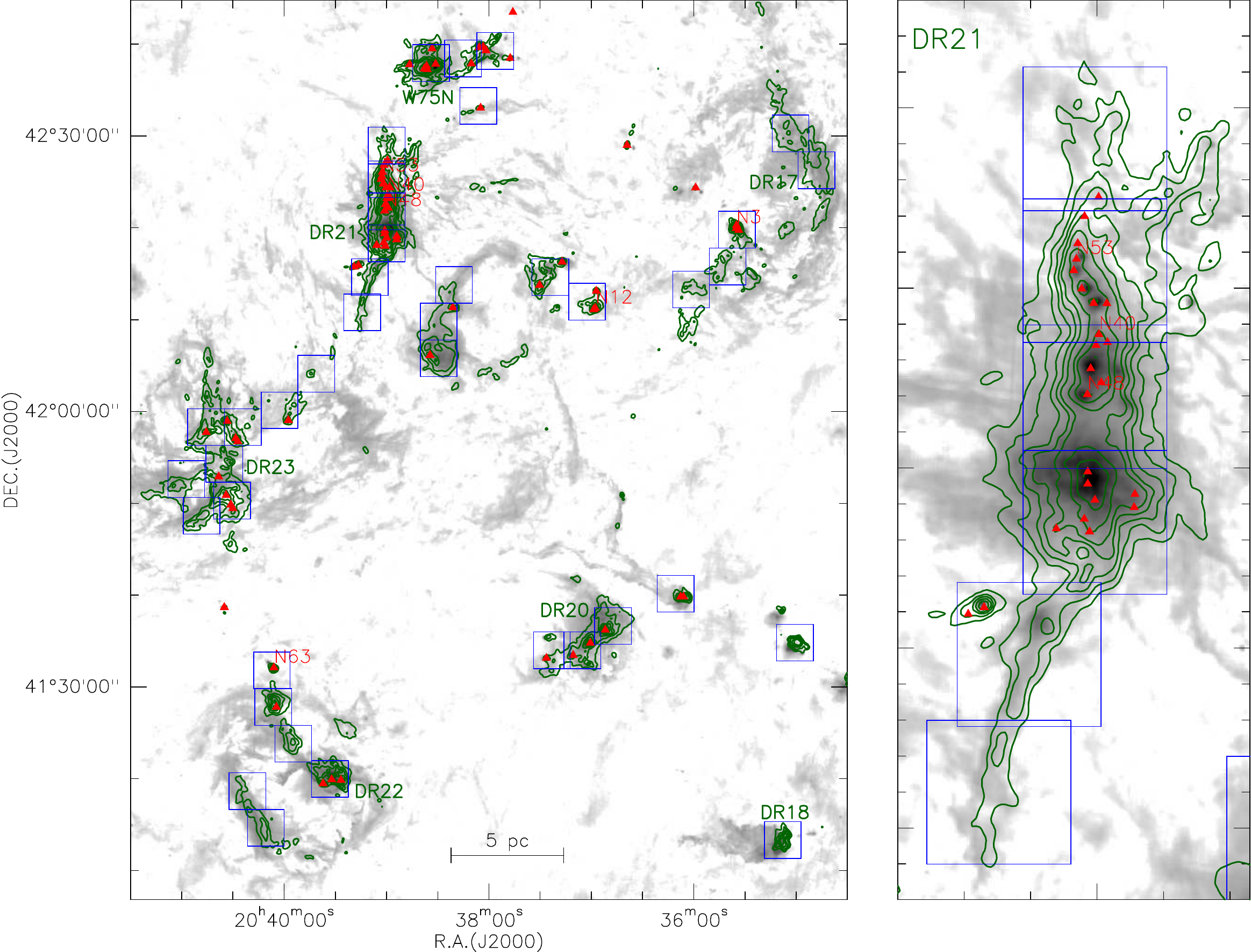}
\caption{Cygnus X as seen in Herschel SPIRE 250\,$\mu$m (green contours)
  and PACS 70\,$\mu$m emission (grayscale, from the HOBYS program,
  \citealt{motte2010}). The red triangles show the mm clumps
  previously identified in $11''$ resolution 1.1 mm single-dish mapping \citep{motte2007}. The
  blue squares indicate the 16\,arcmin$^2$ NOEMA mosaics. To the
  right, a zoom into the DR21 ridge is shown.}
\label{cygx}
\end{figure*}

A complete picture of cloud structure and star formation can be
attained by combining the capabilities of the upgraded Northern
Extended Millimeter Array (NOEMA) with the IRAM 30\,m telescope to
trace all spatial scales. This combination allows sensitive
large-scale mapping of a whole 
spatial resolution with a broad spectral bandpass, sensitively tracing
the continuum emission and a plethora of important spectral
lines. Therefore, we have set up the large Max-Planck-IRAM Observing
Program (MIOP) to map the molecular cloud Cygnus X at high angular
resolution with NOEMA and the IRAM 30\,m telescope: The {\bf C}ygnus
{\bf A}llscale {\bf S}urvey of {\bf C}hemistry {\bf a}nd {\bf
  D}ynamical {\bf E}nvironments (CASCADE). The NOEMA part of the
project is observed in $\sim$520\,h distributed over several
years. Observations of the first regions started during the 2019/2020
winter semester. The complementary IRAM\,30\,m data were observed
between May and August 2020 and will be presented with an overview of
all subregions in Christensen et al.~(in prep.). Furthermore, that
paper will focus on the deuteration of various molecules on large
cloud scales.

With the rich combined NOEMA+30\,m data provided by CASCADE, we will
study a broad range of scientific questions. The key topics to be
addressed relate to the gas flow from large cloud scales to small core
scales. We will investigate potential kinematic signatures of cloud
collapse, and study how signatures may vary for different evolutionary
stages. Furthermore, CASCADE will allows us to scrutinize how
star-forming cores affect their environment, and which feedback
processes from the newly forming stars are dominating. In addition to
this, we will also study the low-mass propulation is such a high-mass
environment.  To be more specific, in the following, we outline a few
topics in more detail.

\begin{figure}[ht] 
\includegraphics[width=0.49\textwidth]{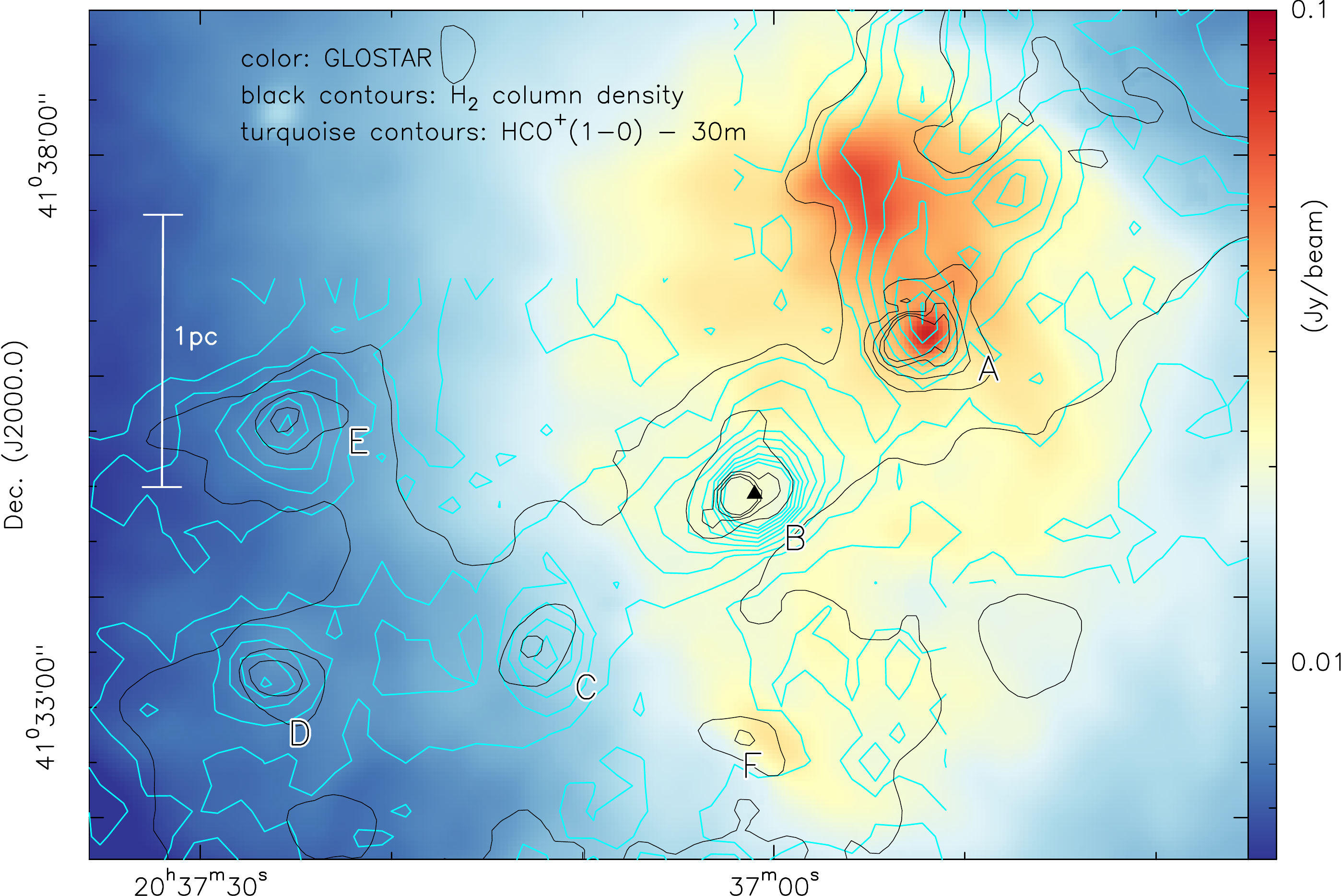}
\includegraphics[width=0.49\textwidth]{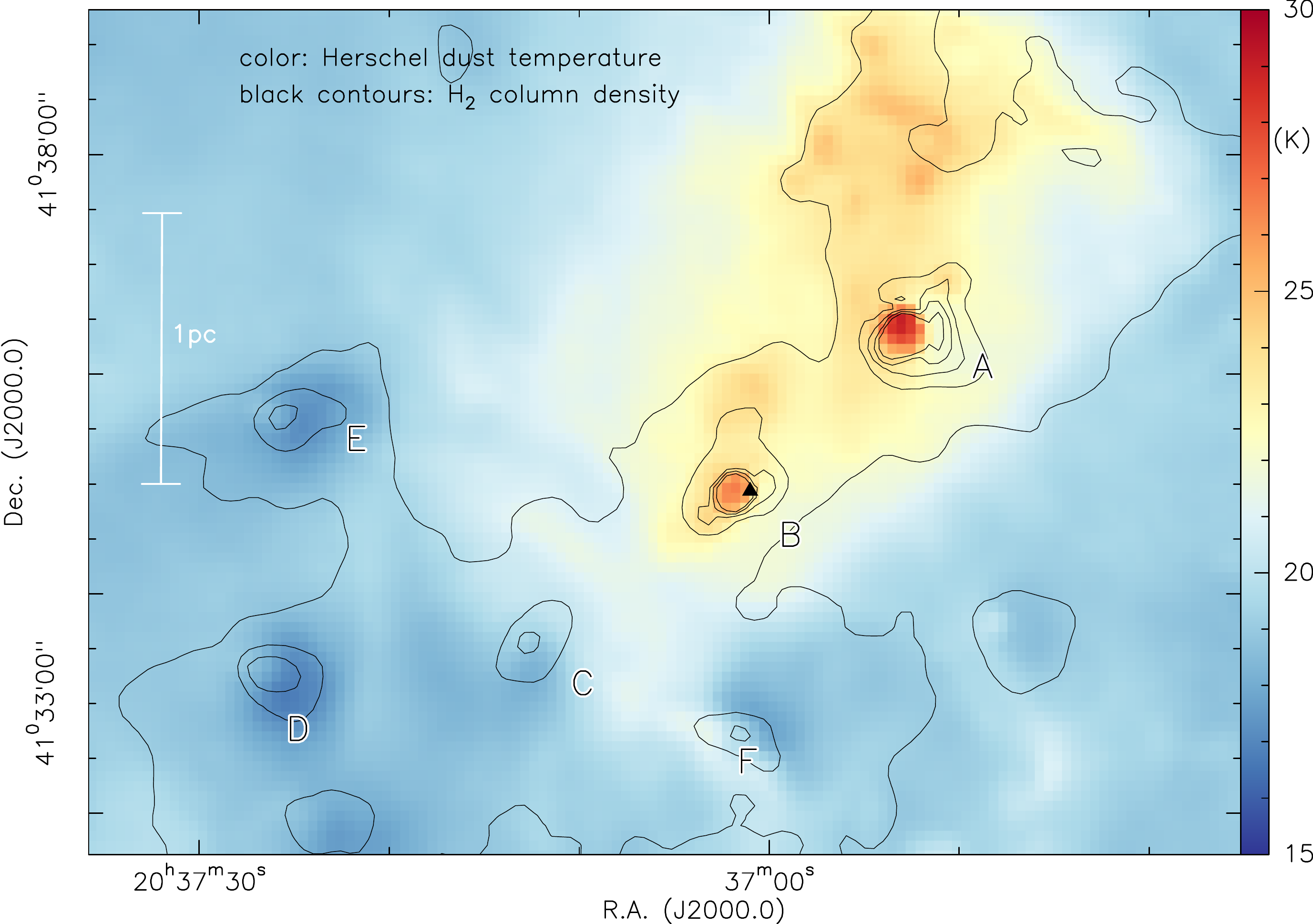}
\caption{Overview of the DR20 region. Top panel: Radio continuum emission from the GLOSTAR survey at
  5.8\,GHz with an angular resolution of $19''$ is shown on the color scale (corresponding to 0.13\,pc;
  \citealt{brunthaler2021}). The turquoise contours present the
  HCO$^+(1-0)$ data obtained with the IRAM 30\,m telescope,
  integrated between $-7$ and 3\,km\,s$^{-1}$. Contour levels start at
  $4\sigma = 3.4$\,K\,km\,s$^{-1}$ and continue in $4\sigma$
  steps. Bottom panel: Dust temperature
  map derived from the Herschel far-infrared continuum data by
  \citet{marsh2017} is show on the color scale. In both panels, the black contours show the H$_2$
  column density map derived by \citet{marsh2017} at contour levels
  between $10^{22}$ and $5\times 10^{22}$\,cm$^{-2}$ (step $1\times 10^{22}$). A linear
  scale-bar is shown at the left. The six main regions A to F are
  marked, and a triangle shows the position of the class II CH$_3$OH
  maser associated with B that has been determined  by \citet{ortiz-leon2021}.} 
\label{dr20}
\end{figure}

\paragraph*{Detailed studies of cloud collapse:} Theoretical models differ in
their predictions for the expected velocity field during the collapse
phase. In the two extreme cases for fast and slow collapse
\citep{larson2003}, much broader line profiles are expected for a
runaway collapse than for quasi-static contraction (e.g.,
\citealt{zhou1992}). While no such broad lines are observed toward
low-mass star-forming regions, the fast collapse might be relevant for
high-mass star-forming cores.  Models of high-mass cluster formation
also predict a large-scale contraction (global collapse) as the
protocluster evolves and molecular gas is funneled from the outer
regions of the core into the center of the cluster
\citep{bonnell2002,bonnell2004,vazquez2019}. The question is how these global
collapse scenarios can be probed.  \citet{vazquez2009,vazquez2019} discussed a
hierarchical global collapse scenario in which the velocity dispersion
on all scales was driven by global infall rather than
turbulence. Observationally, indications have been found of global
infall in low- (e.g., \citealt{walsh2006}) and high-mass star-forming
regions (e.g., \citealt{schneider2010,beuther2020}). With the MIOP
project, we will be able to study the infall kinematics over a broad
range of spatial scales.

\paragraph*{Filaments:} Although the existence of filamentary structures
has been known for some time, Herschel studies revealed the
ubiquity and importance of filaments for the cloud and star
formation process (e.g., \citealt{andre2014}). The general picture
emerging from a wealth of observational and theoretical studies
indicates that molecular clouds often fragment into networks of
filaments, and these filaments further fragment into star-forming
cores. High-mass star formation is often observed at junctions of filaments (so-called hubs; e.g., \citealt{motte2018,kumar2020}). The filaments may act as a funnel through which the gas can
flow, and they link the large-scale cloud to the small-scale
cores. Accretion along filaments has been found from low- to high-mass
star-forming regions (e.g.,
\citealt{tobin2012,kirk2013,peretto2014,hacar2017,hacar2022}). The exact
properties of the filaments are subject of intense debate (e.g.,
\citealt{arzoumanian2011,hacar2013,panopoulou2017,suri2019}), and it
is far from clear whether filamentary structures in high- and low-mass
star-forming regions behave in a self-similar way or if they vary
significantly for the two regimes.
The CASCADE program will allow us to address
several important questions about filaments in high-mass star
formation, e.g., (i) whether we see gas flow motions along filaments, (ii)
whether the gas flows via a network of filaments toward the main
structures, and (iii) we will be able to determine the physical properties of the massive
filaments such as width, velocity dispersion, and mass-to-length ratios.

\paragraph*{Deuterated molecules:} In cold molecular clouds, deuterium enhancement, or deuteration, raises the 
abundance of deuterated molecules relative to their main isotopologs
to values that are orders of magnitude higher than the abundance of D
(relative to H) in the general interstellar medium (where it is $\sim
10^{-5}$; \citealt{oliveira2003}). With the extension of the NOEMA
tuning range to frequencies as low as 71\,GHz, access to several
ground-state transitions of important deuterated species has become
available (e.g., DCN and DCO$^+$; Table \ref{rms}). These observations
will allow studies of the detailed morphology, kinematics, and
deuteration fraction of the individual deuterated species from clump
to core scales and lead to deuteration fractions of the different
species that provide powerful constraints on 
models for the formation route of the deuterated
species. \citet{albertsson2013} modeled the deuteration fraction for a
variety of molecules, initial conditions, timescales, and physical
parameters, which our observations will be able to
constrain. \citet{koertgen2017} performed 3D magnetohydrodynamical
simulations of the deuteration in massive dense prestellar
cores. Interestingly, \citet{pillai2012} found that the fundamental
H$_2$D$^+$ cation in the DR21 ridge in Cygnus X does not follow the
dust continuum. Therefore it is crucial to spatially resolve the
emission of deuterated molecules. Here we wish to understand, for
example, (i) in which density regime N$_2$D$^+$ and N$_2$H$^+$ , for
instance, trace the quiescent gas, (ii) which other deuterated species
behave similarly (e.g., DCO$^+$, DCN, or DNC), and (iii) we aim to
determine the influence of the deeply embedded low-luminosity young
stellar objects (YSOs) on the deuteration.

\section{The nearby  Cygnus X region: A star formation powerhouse}

The target region for CASCADE is the well-studied northern
hemisphere Cygnus X complex (e.g., \citealt{reipurth2008}, Fig.~\ref{cygx}). This
relatively nearby ($\sim 1.4$\,kpc, \citealt{rygl2012}) massive and luminous
region (e.g., \citealt{kumar2007,beerer2010}) is an excellent hunting
ground in which to study various phases of massive star formation.  Its
associated giant molecular cloud complex harbors a rich collection of objects in 
various star formation stages: 
dense, dusty, hot cores with embedded protostars (e.g.,
\citealt{schneider2006,motte2007}). This and the presence of several
OB associations, the results of many million years of recent star formation, and a super-bubble driven by stellar winds of the O stars and the famous Cygnus loop supernova remnant
(also known as the~Diamond ring) give testament to intense and widespread star
formation that occurred over (at least) the past few million
years. More recently, mm and centimeter (cm) wavelength observations have resulted
in a wealth of data on the dust, ionized, and molecular gas
emission in this region (e.g.,
\citealt{taylor2003,motte2007,motte2010,molinari2010,cao2019,cao2021,cao2022,li2021,vanderwalt2021}). Taken
together, these substantial complementary data form a rich background
that now allows placing all spatial scales into context and following
the gas flow from the large molecular clouds to the small cores. The
Cygnus region is thus one of the few relatively nearby regions in which
many phenomena can be studied in detail.

\begin{figure}[htb]
\includegraphics[width=0.49\textwidth]{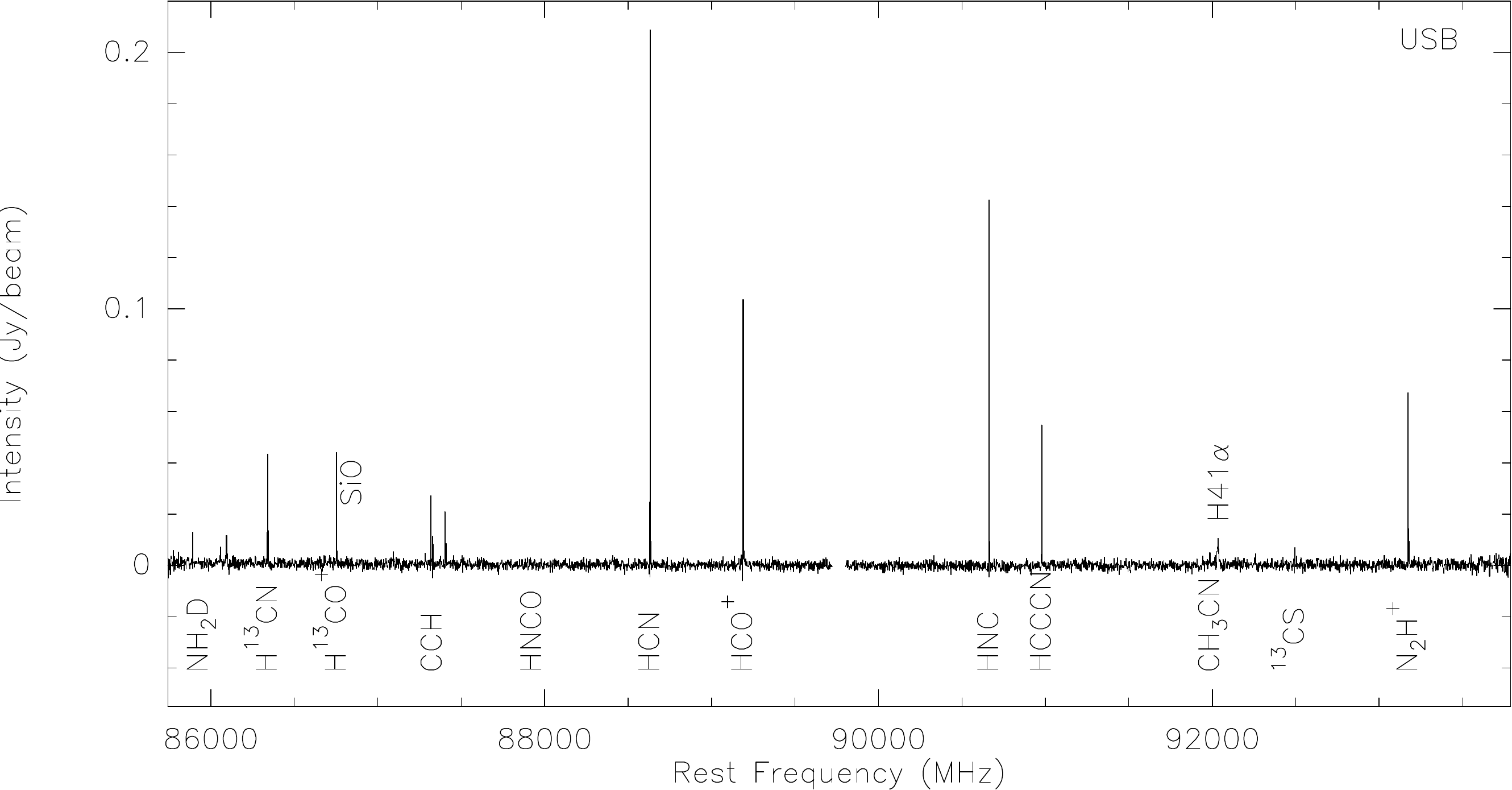}
\includegraphics[width=0.49\textwidth]{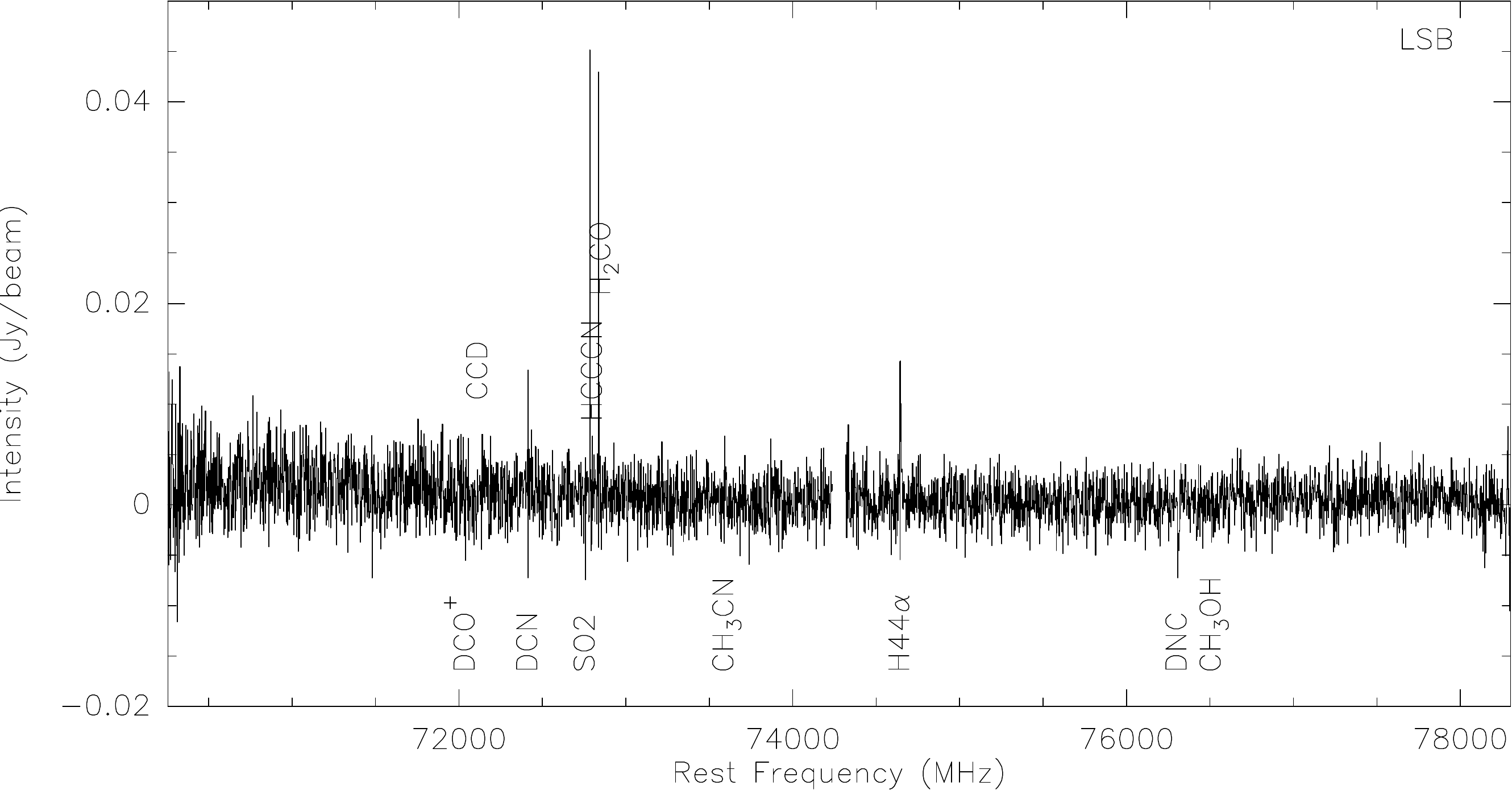}
\caption{NOEMA+30\,m wide-band spectra toward the main continuum peak position A.}
\label{spectra} 
\end{figure} 

\section{The Max Planck IRAM Observatory program (MIOP) on star formation in Cygnus X}
\label{miop}

Figure \ref{cygx} gives an overview of the Cygnus X region observed
with Herschel at the far-infrared wavelength (250\,$\mu$m). While the
entire complex covers several square degrees on the sky, the
high-column density regions visible in the Herschel data are
constrained to comparably smaller areas. The CASCADE program targets
all the important subregions with 40 mosaics covering 16\,arcmin$^2$
each (each blue square in Fig.~\ref{cygx} marks one mosaic). Several
individual and adjacent mosaic fields are then combined into larger
mosaics. For example, the prominent DR21 region shown in the right
panel of Fig.~\ref{cygx} comprises six mosaics.

Our selected spectral setup in the 3.6\,mm wavelength band of
$\sim$8\,GHz width in the upper and lower sideband 
covers
crucial ground-state lines from N$_2$H$^+$, HCN, HNC, and H$_2$CO, for example,
as well as unique lines from deuterated molecules. Figure
\ref{spectra} shows example spectra taken toward the main continuum
emission peak in the DR20 region with the broadband correlator unit, which has a spectral resolution of 2.0\,MHz, corresponding to a
velocity resolution of $\sim$7.3\,km\,s$^{-1}$ at the central nominal
frequency of $\sim$82.028\,GHz (Table \ref{rms}). Important lines are
marked. Most of these lines are also covered by additional
high-resolution correlator units to also resolve them at a high spectral resolution of 60\,kHz, corresponding to a velocity resolution
between 0.26 and 0.19\,km\,s$^{-1}$ at the bottom and top end of the
bandpass (Fig.~\ref{spectra}). Table \ref{lines} presents all spectral
units, covered frequencies, and specific associated lines. These lines
cover a broad range of 
energies above the ground state, critical densities,
optical depths, and chemical properties. This allows us to probe the kinematics and the physical and chemical conditions of the dense
regions on 5000\,au scales in detail.

\begin{figure}[ht] 
\includegraphics[width=0.49\textwidth]{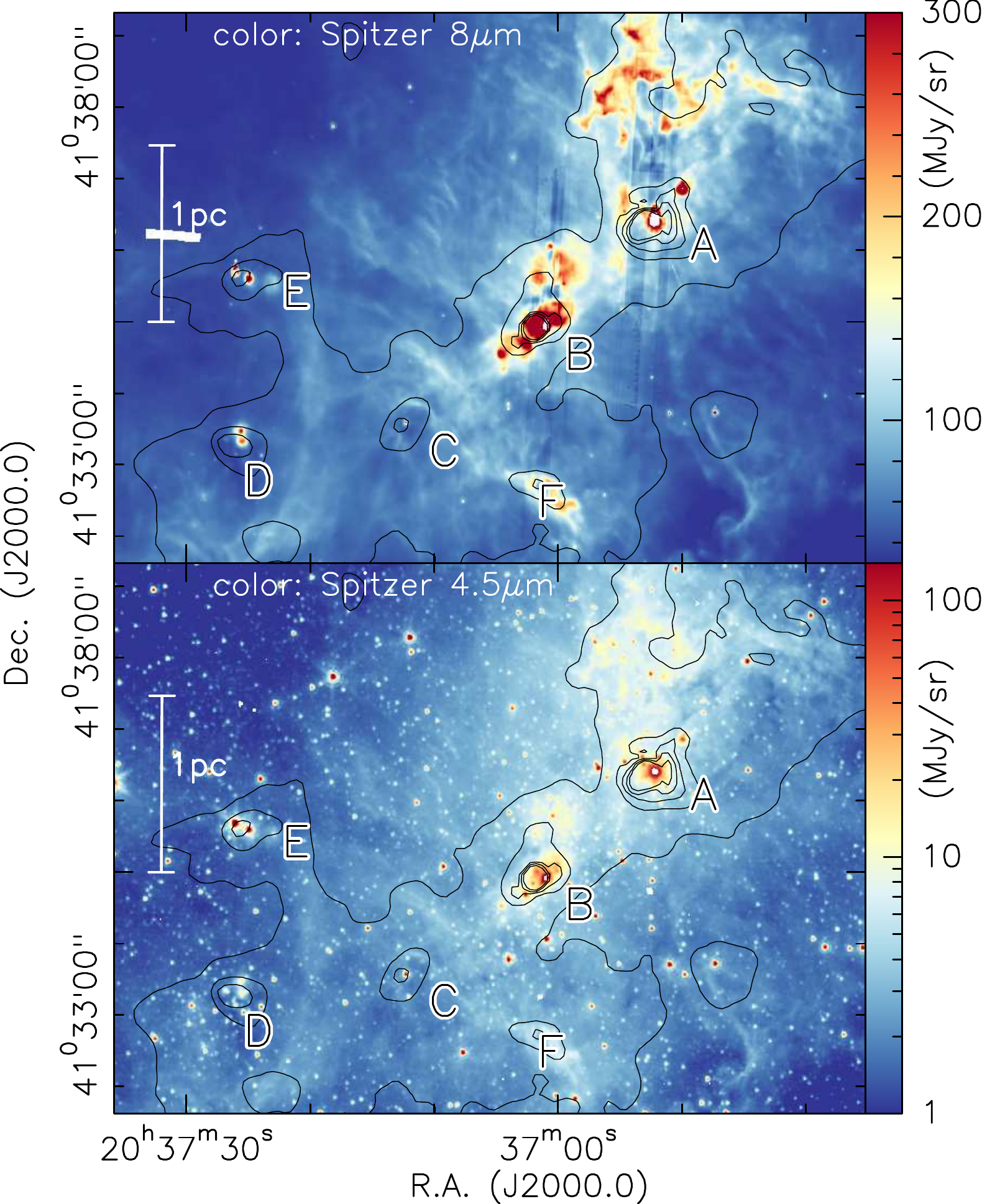}
\caption{ Spitzer 8\,$\mu$m and
  4.5\,$\mu$m data on the color scale. The black contours again show the
  H$_2$ column density map in contour levels between $10^{22}$ and
  $5\times 10^{22}$\,cm$^{-2}$ (step $1\times 10^{22}$\,cm$^{-2}$).}
\label{dr20_spitzer}
\end{figure}

Each individual mosaic consists of 78 pointings, and we observe
all regions in the C and D configurations. This NOEMA part of CASCADE
is planned to take roughly 520\,h. To complement the missing short
spacings, we furthermore applied for open time at the IRAM 30\,m
telescope. Within $\sim$80\,h of observing time, we
already observed all fields with the IRAM 30\,m telescope in 2020. The entire IRAM 30\,m
dataset is presented in a forthcoming paper by Christensen et al..

The following case study focuses on the Cygnus X subregion DR20, which was first
discussed by \citet{downes1966}. It is a strong radio and mm source
that was investigated in various studies (e.g.,
\citealt{schneider2006,motte2007}) and is also part of the recent
GLOSTAR Galactic plane survey \citep{brunthaler2021}. Figure
\ref{dr20} shows an overview of the complex based on Herschel dust
continuum and GLOSTAR radio continuum data
\citep{marsh2017,brunthaler2021}. DR20 exhibits an almost filamentary
structure from the northwest toward the southeast, and it combines sources in a
range of evolutionary stages, from more evolved (ultracompact) H{\sc
  ii} regions (clumps A and B in Fig.~\ref{dr20}) to the younger and
colder clumps C to F. This variety of evolutionary stages makes DR20
an ideal template for a first exploration of the potential of the
CASCADE project.

The DR20 region, the focus of  this paper, was observed with three
mosaic tiles, covering star-forming regions in different evolutionary
stages (Fig.~\ref{dr20}). All NOEMA and IRAM 30\,m data have been
taken, and we present the full combined dataset. This allows us to also
discuss the observation and data reduction challenges
associated with the project.

\begin{figure*}[h]
\includegraphics[width=0.9\textwidth]{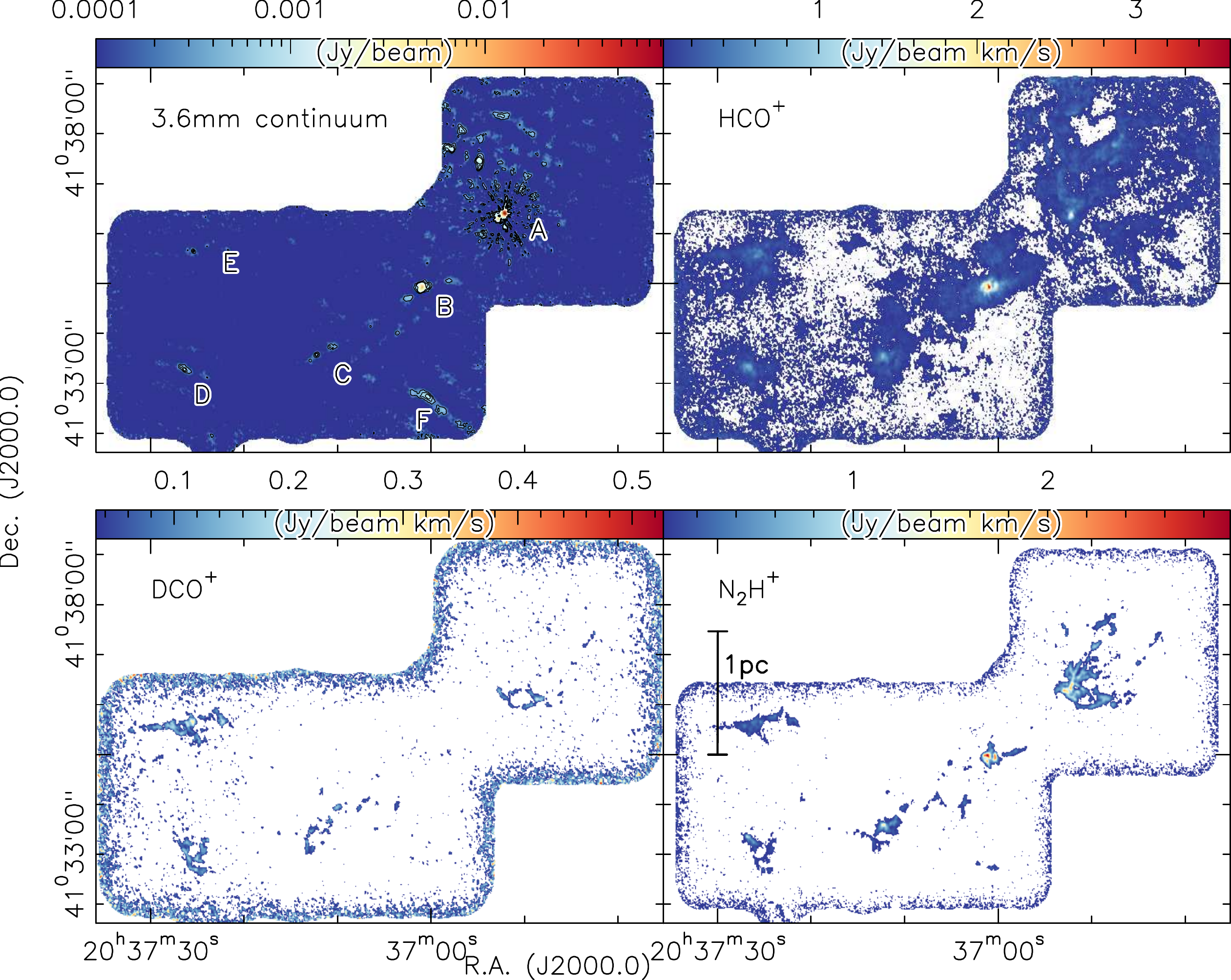}
\caption{Millimeter continuum and selected integrated intensity
  images. The upper left panel shows the 3.6 mm continuum emission
  (NOEMA only), and the other panels show integrated intensity images
  for the species marked in each panel (NOEMA+30\,m). For the latter,
  the integration ranges are always from $-7$ to 3\,km\,s$^{-1}$. The
  data were clipped below the $4\sigma$ level for each species (Table
  \ref{rms}). The bottom right panel also shows a linear
  scale-bar. The six main regions A to F are labeled in the continuum
  panel. The continuum image also shows the lower continuum contour
  levels in 6$\sigma$ steps of 0.3\,mJy\,beam$^{-1}$ up to
  0.9\,mJy\,beam$^{-1}$. Additional maps for other species are shown
  in Figures \ref{mom0_2} and \ref{mom0_3} in the appendix.}
\label{mom0} 
\end{figure*} 

\section{Observations and data reduction}
\label{data}

The three mosaic tiles of the DR20 region were observed individually
with NOEMA in the C and D configurations between 2019 December 15
and 2020 May 3, with typically ten antennas in the array (baselines between $\sim$15 and $\sim$365\,m). Each mosaic
tile consists of 78 pointings. The phase reference centers (in
J2000.0) of the three fields are
R.A.~20:36:47.570 Dec.~+41:36:50.40, R.A.~20:37:05.437 Dec.~+41:34:10.40, and R.A.~20:37:23.303 Dec.~+41:34:10.40.

The $\varv_{\rm{LSR}}$ for all three fields is the same with
$-2.0$\,km\,s$^{-1}$. The bandpass and flux calibration was conducted
with the quasar 3C454.3 and with MWC349, respectively. Gain
calibration was performed via regularly interleaved observations of
the quasars 2005+403 and 2120+445.

A technical challenge of this MIOP project is achieving the desired
homogeneous sensitivity distribution throughout the 40 mosaic tiles.
A complete track that observes one full mosaic of 78 pointings with three coverages, which is required to reach the targeted mosaic sensitivity of
$\sim$11\,mJy\,beam$^{-1}$ in 150\,kHz channels,~takes about 9\,h of
telescope time. It is therefore important to protect against possibly
varying weather conditions during individual tracks, and
against technical failures that could unexpectedly interrupt an observing track when only part of the complete mosaic is
observed. To reach this goal, a suite of dedicated observing
procedures was developed that easily allow to add additional data to
individual parts of the mosaic, when needed.

The calibration and imaging was performed with the {\sc clic} and {\sc
  mapping} software within the {\sc gildas}
package\footnote{http://www.iram.fr/IRAMFR/GILDAS}. The continuum
dataset was created using only the line-free parts of the
spectrum. Excluding the band edges and the lines, we still have a very
broad bandpass of $\sim$15.51\,GHz for our final continuum dataset.
To create the final continuum table, we first averaged over 20\,MHz bins to properly rescale the uv coordinates to the mean frequency of each bin. The {\sc uv\_cont} task then created the final continuum table, taking the spectral slope of the data into account. We
applied natural weighting during the imaging process for optimal
imaging quality in these large mosaics. For the 3.6\,mm continuum
data, this resulted in synthesized beams of $3.45''\times 2.77''$
(P.A.~21\,deg), corresponding to an approximate linear spatial
resolution of $\sim$4400\,au. The achieved $1\sigma$ rms noise level of the
continuum data is $\sim$50\,$\mu$Jy\,beam$^{-1}$. The absolute
flux scale is estimated to be correct within 10\%.

For the spectral line data, we obtained complementary single-dish
observations with the IRAM 30\,m telescope to compensate for the
missing short spacings. These 30\,m observations were conducted in
on-the-fly mode, typically achieving rms values of $\sim$0.1\,K
($T_{\rm{mb}}$).  The IRAM 30\,m data of the entire MIOP program
toward Cygnus X are presented in Christensen et al.~(in prep.).

We then combined the NOEMA and 30\,m data in the imaging process with
the task {\sc uv\_short}. All spectral line results presented in this
paper are based on the combined NOEMA C+D array plus IRAM 30\,m
observations. Because the spectral resolution of the 30\,m data is
slightly lower, we used a homogeneous velocity resolution of
0.8\,km\,s$^{-1}$. Only the broad hydrogen recombination lines were
imaged at a lower velocity resolution of 3\,km\,s$^{-1}$.  Again,
natural weighting was applied, and the final $1\sigma$ rms and
synthesized beam values are presented in Table \ref{rms}. The highest
angular resolution is achieved for the N$_2$H$^+(1-0)$ line at
93.174\,GHz with $3.04''\times 2.28''$ or $\sim$3700\,au.

\begin{table}[htb]
\caption{Continuum and spectral line parameters.}
\begin{tabular}{lrrr}
  \hline \hline
line & freq. & $1\sigma$ & beam \\
& (GHz) & $\left(\frac{\rm mJy}{\rm beam}\right)$ \\
\hline
continuum & 82.028 & 0.05 & $3.45''\times 2.77''$ \\
DCO$^+$(1--0) & 72.039 & 10 & $4.11''\times 3.12''$ \\
CCD(1--0) & 72.108 & 10 & $4.11''\times 3.12''$ \\
DCN(1--0) & 72.415 & 9 & $4.09''\times 3.11''$\\
SO$_2(6_{0,6}-5_{1,5})$ & 72.758 & 9 & $4.07''\times 3.09''$\\
HCCCN(8--7) & 72.784 & 8 & $4.07''\times 3.09''$ \\
H$_2$CO$(1_{0,1}-0_{0,0})$ & 72.838 & 8 & $4.07''\times 3.09''$ \\
CH$_3$CN$(4_k-3_k)$ & 73.590 & 8 & $4.03''\times 3.07''$ \\
H44$\alpha$ & 74.645 & 4.5 & $3.91''\times 3.00''$\\
DNC(1--0) & 76.306 & 7 & $3.85''\times 2.93''$ \\
CH$_3$OH$(5_{0,5}-4_{1,3})$E & 76.510 & 7 & $3.84''\times 2.93''$ \\
NH$_2$D$(1_{1,1}-1_{0,1})$ & 85.926 & 6 & $3.31''\times 2.49''$ \\
H$^{13}$CN(1--0) & 86.340 & 7 & $3.29''\times 2.47''$ \\
H$^{13}$CO$^+$(1--0) & 86.754 & 6 & $3.28''\times 2.47''$ \\
SiO(2--1) & 86.847 & 6 & $3.27''\times 2.46''$  \\
HN$^{13}$C(1--0) & 87.091 & 7 & $3.27''\times 2.46''$  \\
CCH(1--0) & 87.329 & 6 & $3.26''\times 2.45''$ \\
HNCO$(4_{0,4}-3_{0,3})$ & 87.925 & 6 & $3.23''\times 2.44''$ \\
HCN(1--0) & 88.632 & 7 & $3.21''\times 2.42''$ \\
HCO$^+$(1--0) & 89.189 & 6 & $3.19''\times 2.41''$ \\
HNC(1--0) & 90.664 & 7 & $3.13''\times 2.33''$ \\
HCCCN(10--9) & 90.979 & 7 & $3.12''\times 2.32''$ \\
CH$_3$CN$(5_k-4_k)$ & 91.987 & 7 & $3.08''\times 2.30''$\\
H41$\alpha$ & 92.034 & 4.5 & $3.08''\times 2.30''$\\
$^{13}$CS(2--1) & 92.494 & 7 & $3.06''\times 2.29''$\\
N$_2$H$^+$(1--0) & 93.174 & 8 & $3.04''\times 2.28''$\\
\hline \hline
\end{tabular}
~\\
Notes: The $1\sigma$ rms values are for all lines with a resolution of 0.8\,km\,s$^{-1}$ , except of the hydrogen recombination lines, which have a resolution of 3\,km\,s$^{-1}$ . The continuum rms is for the whole bandpass, excluding the strong spectral lines.
\label{rms}
\end{table}

\begin{figure}[htb]
\includegraphics[width=0.49\textwidth]{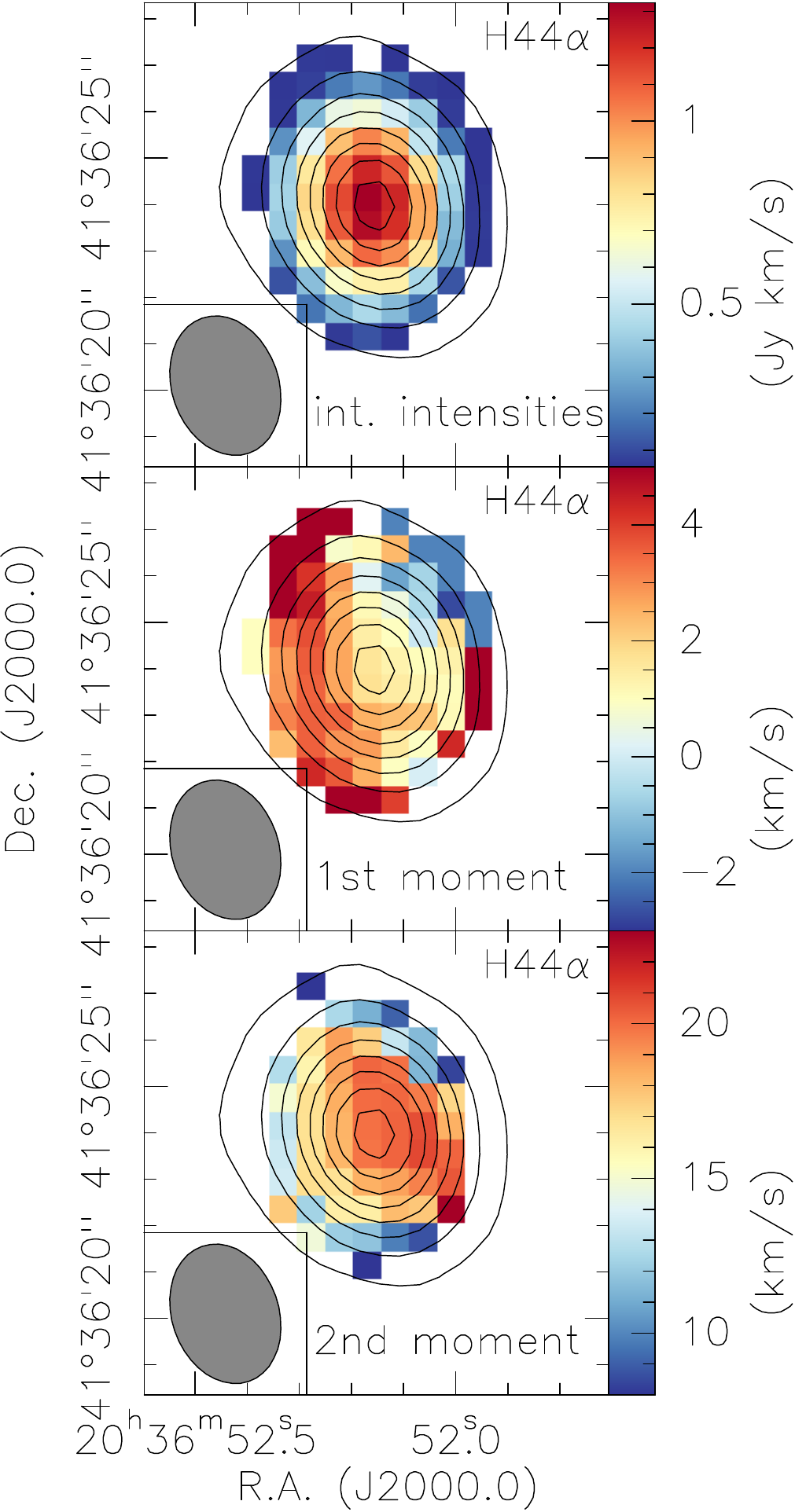}
\caption{Integrated intensities (top), first- (middle) and second
-moment (bottom) maps of the  H$44\alpha$ radio recombination line are
  shown in color scale.  panel. The integration ranges are always from
  -20 to 20\,km\,s$^{-1}$. The data were clipped below the $4\sigma$
  level (Table \ref{rms}). The contours show the 3.6\,mm continuum
  data on contour levels from 8 to 78\,mJy\,beam$^{-1}$ (step
  10\,mJy\,beam$^{-1}$). The beam is shown in the bottom left corner of each
  panel.}
\label{rrl} 
\end{figure} 

\section{Results}
\label{results}

\begin{figure*}[htb]
\includegraphics[width=0.99\textwidth]{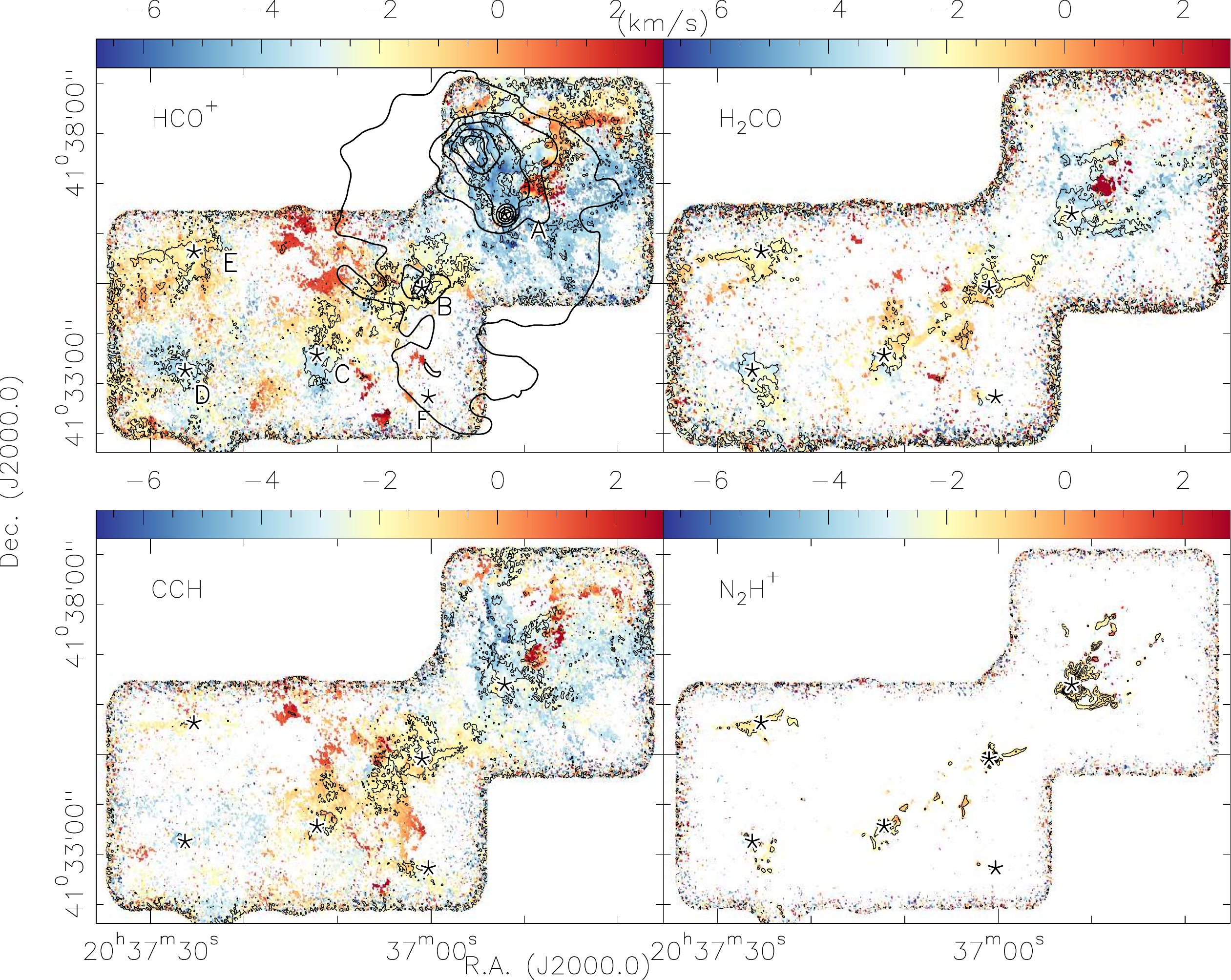}
\caption{First-moment maps. The colour scale always shows the first
  moment for the labeled species. The black contours
  outline the integrated emission of the same species contoured always
  from 0.1 to 2.9\,Jy\,beam$^{-1}$\,km\,s$^{-1}$ in steps of
  0.3\,Jy\,beam$^{-1}$\,km\,s$^{-1}$. The thick black
  contours in the top left panel show the GLOSTAR 5.8\,GHz continuum
  data between 0.02 and 0.12\,Jy\,beam$^{-1}$ in steps of
  0.01\,Jy\,beam$^{-1}$. The 3.6\,mm continuum peak positions are marked with five-point stars.}
\label{mom1} 
\end{figure*} 

Figures \ref{dr20} and \ref{dr20_spitzer} present overviews of the
DR20 region by showing the GLOSTAR 5.8\,GHz radio continuum emission
\citep{brunthaler2021}, two Spitzer mid-infrared bands (8 and 4.5\,$\mu$m), and the
integrated emission of the HCO$^+(1-0)$ line obtained with the IRAM
30\,m telescope. Furthermore, the H$_2$ column densities and
the dust temperatures derived by \citet{marsh2017} from the HIGAL data
\citep{molinari2016b} are shown. As mentioned above, 
a gradient in star formation activity over the region can be identified in which the
western part already hosts an H{\sc ii} region (near A) that is shown by the radio
and strong mid-infrared emission. Region B also shows elevated temperatures, strong compact molecular emission, and a class II CH$_3$OH maser found by \citet{ortiz-leon2021}. Class II CH$_3$OH masers are only found toward high-mass YSOs that are mostly in the pre-ultracompact H{\sc ii} region stage \citep{minier2003}.

The eastern part is comparatively more
quiescent at radio and mid-infrared wavelengths. This gradient is also
reflected in the dust temperature maps in which we find elevated
temperatures of about 30\,K in the already active regions A and B,
whereas the other four regions rather exhibit dips in the temperature
map below 20\,K. The HCO$^+(1-0)$ emission agrees well spatially with
the general H$_2$ column density structure derived from the
far-infrared HIGAL data. The total gas mass estimated from the H$_2$
column density map in the area shown in Fig.~\ref{dr20} is
$\sim$2500\,M$_{\odot}$.

The different parts of the DR20 region follow an almost filamentary
structure in northwest-southeast direction along four main clumps
labeled A to D in Fig.~\ref{dr20}. Adjacent to this main
filament, we find two more clumps E and F, also labeled in Figure
\ref{dr20}. The faintest clump F can barely be identified in the
HCO$^+$ data, but is also emitting at mid-infrared wavelengths. The
nearest-neighbor separations for these main clumps are about 1\,pc (Table \ref{nearest}). 
Very interestingly, the classical
Jeans length for a 20\,K cloud at a mean density of 750\,cm$^{-3}$ is
roughly 1\,pc as well (e.g., \citealt{stahler2005}). A detailed analysis of the filamentary substructure and associated kinematic features will be presented in Sawczuck et al.~(in prep.). That analysis will be based on a DISPERSE filament identification \citep{sousbie2011a,sousbie2011b} and the subsequent analysis and physical interpretation of the filamentary structures.

\begin{table}[htb]
\caption{Nearest-neighbor separations of the main clumps}
\begin{tabular}{lr}
  \hline \hline
clumps & sep. \\
  & (pc) \\
\hline
A-B & 0.88 \\
B-C & 1.11 \\
C-D & 1.08 \\
D-E & 0.96 \\
C-F & 0.97 \\
\hline \hline
\end{tabular}
\label{nearest}
\end{table}

\subsection{General gas morphologies}

\begin{figure*}[htb]
\includegraphics[width=0.99\textwidth]{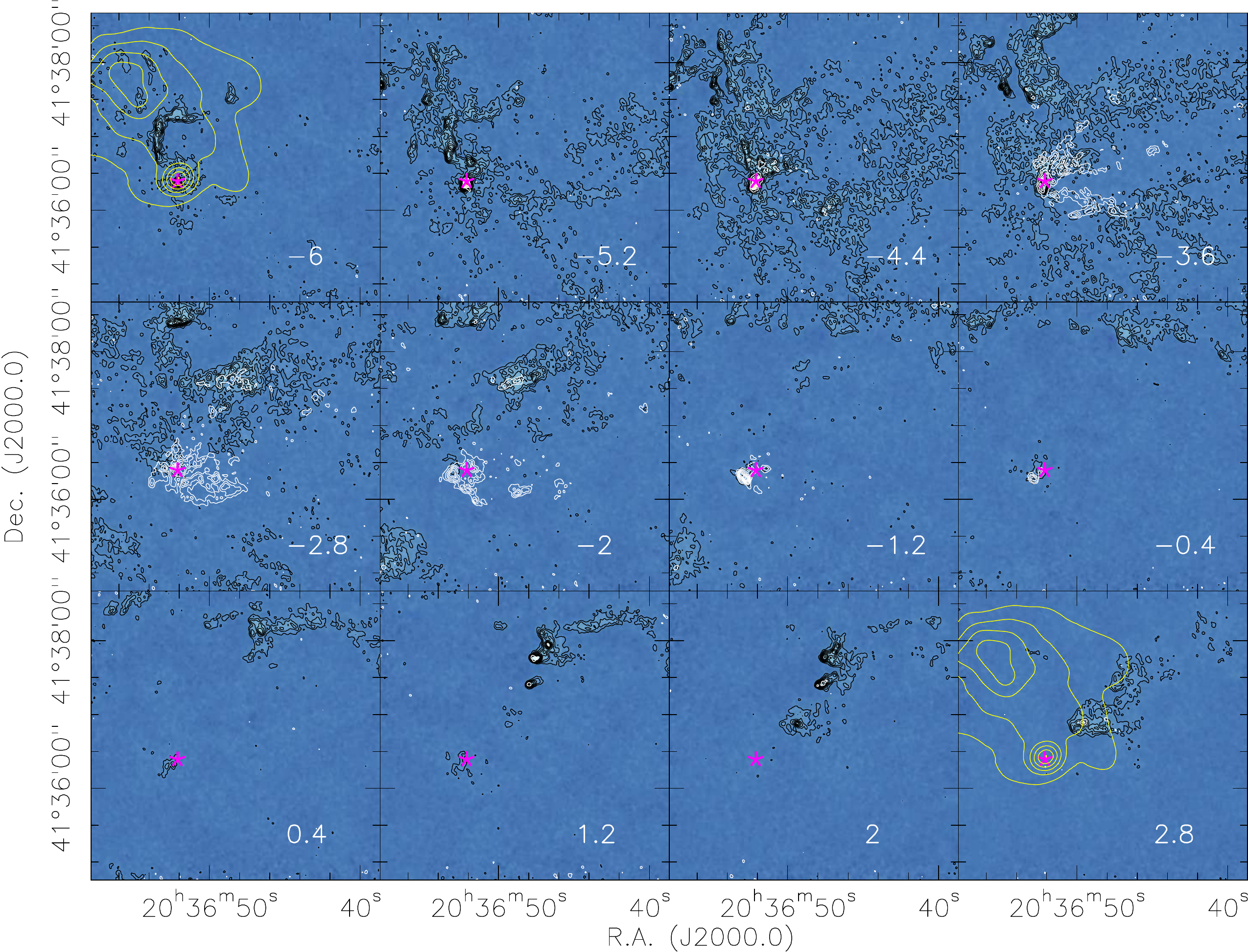}
\caption{Channel maps of the western third of DR20 around region A in the
  HCO$^+(1-0)$ and H$^{13}$CO$^+(1-0)$ emission lines. The colored and
  black contours show the HCO$^+(1-0)$ emission, whereas the white
  contours present the H$^{13}$CO$^+(1-0)$ emission. Contour levels
  are always in 5$\sigma$ steps, and the central velocity is marked in
  each panel. The magenta five-point stars marks the position of the main
  continuum peak (Fig.~\ref{mom0}). The yellow contours in the first
  and last panel outline the GLOSTAR cm continuum data from 0.03 to
  0.12\,Jy\,beam$^{-1}$ in steps of 0.01\,Jy\,beam$^{-1}$.}
\label{hco+_channel} 
\end{figure*}

Figure \ref{mom0} and Figures \ref{mom0_2} and \ref{mom0_3} in the
appendix present the 3.6\,mm NOEMA-only continuum and the integrated
intensity maps of the combined NOEMA+30\,m line data. These images
show the richness of the data obtained in this large program, which
combines the capabilities of the two IRAM facilities, NOEMA and the
30\,m telescope. The continuum image still exhibits artifacts and
increased noise around the strong emission sources that is caused by
insufficient deconvolution because of the missing short spacings for
the continuum. Except for the increased noise at the map and mosaic
edges, the combined spectral line data are free from such artifacts
and recover the large-scale and the small-scale emission. This allows
us for the first time to study the gas properties from cloud scales
down to scales of individual protostellar cores. The DR20 region is an
ideal early example of the survey because it combines several aspects
to be studied by this MIOP: It covers a range of evolutionary stages
from an evolved H{\sc ii} region in the west (subregion A), close to
an already ignited high-mass protostellar object whose mid-IR
radiation powers a class II CH$_3$OH maser, to more pristine and
younger regions farther east (Figs.~\ref{dr20} and
\ref{dr20_spitzer}). The region is observable in many spectral lines,
and it harbors several structures that are connected and
interacting. We discuss all these points in the following.

Because the different molecules trace different parts of the gas, our
data allow us to investigate more diffuse gas components and
dense components. For example, the HCO$^+$, HCN, or CCH emission traces
more widespread and diffuse gas components, whereas other transitions,such as those from H$^{13}$CO$^+$ or N$_2$H$^+$ , trace the denser parts
of the star-forming regions. Even denser regions can then be studied by
CH$_3$CN or HCCCN. The emergence of deuterated species such as DCO$^+$, DCN, DNC, or
NH$_2$D does not depend on density alone, but also on chemistry,
and the isotopologs trace only specific parts of the gas (see section
\ref{deuteration}). For example, subregion F in the south of DR20 is
clearly detected in the mm continuum, but only barely visible in the HCO$^+$ line
emission. In contrast to the latter, region F is again well detected
in spectral lines of H$_2$CO, CCH, or HNC, but it is far fainter in HCN. Hence,
we identify clear differences in the chemistry throughout the entire DR20
region. Dedicated chemical investigations of the entire CASCADE data are planned as forthcoming studies.

Furthermore, we also cover two hydrogen
radio recombination lines (H41$\alpha$ and H44$\alpha$) that are
typically only detected toward the strongest free-free continuum
sources within a field. In the case of DR20, they are detected toward
the strong cm continuum source in the west (subregion A,
Fig.~\ref{dr20}), which at the same time is also the strongest 3.6\,mm
source in the DR20 field (Fig.~\ref{mom0}). Fig.~\ref{rrl} shows the
corresponding integrated intensities and first- and second-moment maps. We
return to this in section \ref{kinematics}.

\begin{table*}[h]
  \caption{Core parameters from the 3.6\,mm continuum data.}
  \begin{tabular}{lrrrrrrrrrr}
    \hline \hline
    \# & $S_{\rm{peak}}$ & $S_{\rm{int}}^a$ & $S_{\rm{peak-D}}^b$ & $S_{\rm{int-D}}^b$ & $S_{\rm{peak-B}}^c$ & $S_{\rm{int-B}}^c$ & $N^g_{\rm{gas}}$ & $M^g_{\rm{gas}}$ & $N^g_{\rm{gas}}$@50K & $M^g_{\rm{gas}}$@50K \\
    & ($\frac{\rm{mJy}}{\rm{beam}}$) & (mJy) & ($\frac{\rm{mJy}}{\rm{beam}}$) & (mJy) & ($\frac{\rm{mJy}}{\rm{beam}}$) & (mJy) & ($10^{24}$cm$^{-2}$) & (M$_{\odot}$) & ($10^{24}$cm$^{-2}$) & (M$_{\odot}$) \\
    \hline
    A & 82.9 & 108.0 & 21.3 & 46.4 & 57.9 & 82.7 & 5.8-15.7$^d$ & 116.9-208.3$^d$ & 3.3-8.9$^d$ & 65.9-117.4$^d$ \\
    B & 9.8 & 36.0 & 0 & 20.3 & & & 2.9$^e$ & 55.2-97.9$^e$ & 1.5$^e$ & 28.8-51.1$^e$ \\
    C$^f$ & 1.3 & 2.0 & & & & & 0.5 & 7.5 \\
    D$^f$ & 1.5 & 4.8 & & & & & 0.6 & 18.1 \\
    E$^f$ & 1.0 & 1.0 & & & & & 0.4 & 3.8 \\
    F$^f$ & 1.1 & 12.0 & & & & & 0.5 & 45.2 \\
    \hline \hline
  \end{tabular}
  ~\\
  $^a$ Integrated within 6$\sigma$ contours.\\
  $^b$ Peak and integrated flux densities, free-free subtracted based on the GLOSTAR 5.8\,GHz D-array data within the $19''$ beam.\\
  $^c$ Peak and integrated flux densities, free-free subtracted based on the GLOSTAR 5.8\,GHz B-array integrated flux density.\\
  $^d$ Column density and mass ranges based on the two free-free subtracted flux densities.\\
  $^e$ Column density without free-free correction. Mass range based on D-array free-free subtracted and no free-free subtraction.\\
  $^f$ no free-free contribution for these cores.\\
  $^g$ $N_{\rm{gas}}$ and $M_{\rm{gas}}$ are calculated with the temperatures discussed in the text, whereas  $N_{\rm{gas}}$ at 50K and $M_{\rm{gas}}$ at 50K are estimated assuming 50\,K for regions A and B.
  \label{core_parameters}
\end{table*}

\begin{figure*}[htb]
\includegraphics[width=0.99\textwidth]{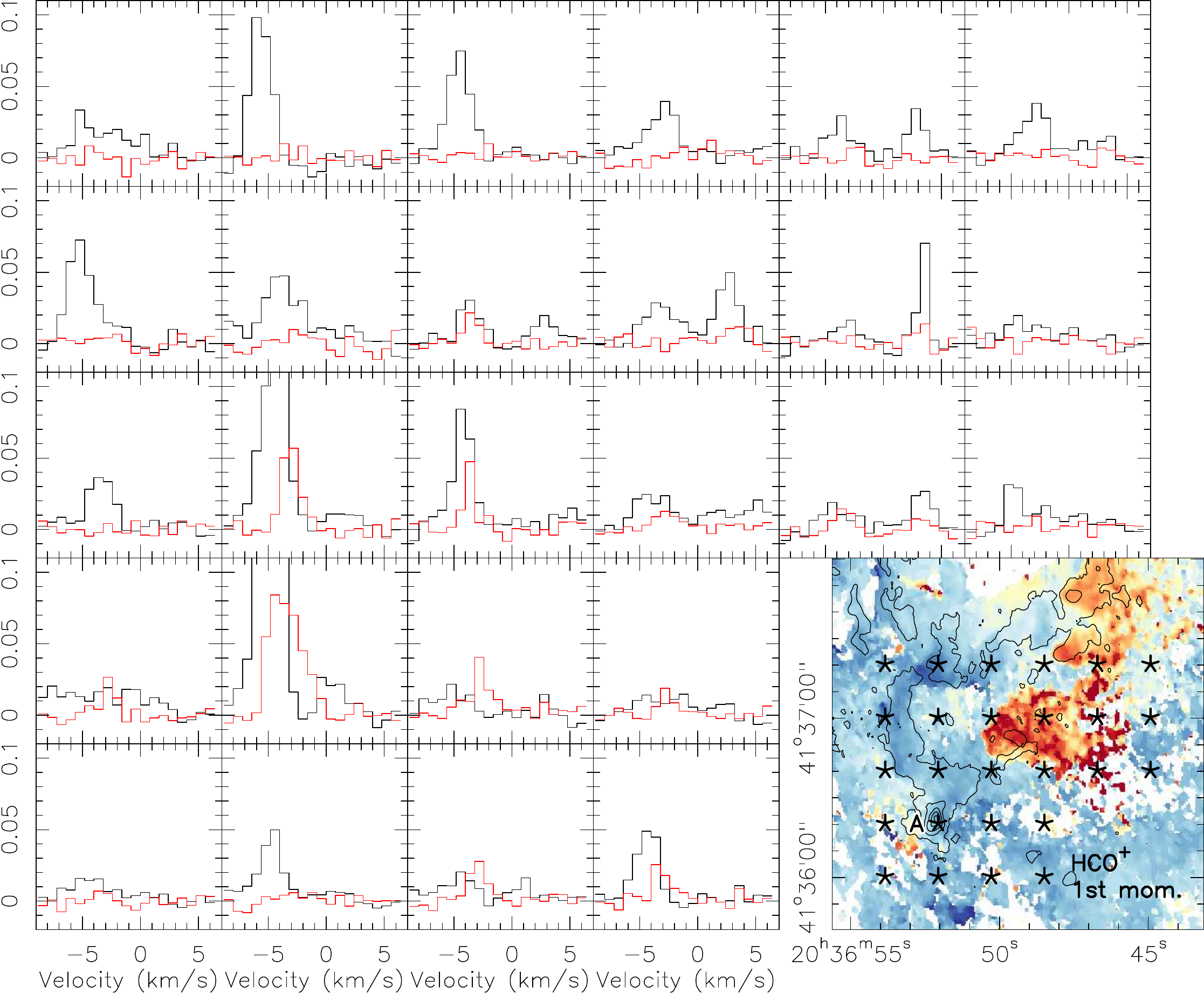}
\caption{HCO$^+(1-0)$ and H$^{13}$CO$^+(1-0)$ spectra in black and
  red, respectively, toward the northwestern region A of DR20. The y-axis of the spectra shows the flux densities in units of Jy\,beam$^{-1}$. The
  color-scale in the bottom right panel shows the corresponding
  HCO$^+$ first-moment map, in which the stars mark the positions of the
  presented spectra. The contours show the HCO$^+(1-0)$ integrated
  intensity map (from -7 to 3\,km\,s$^{-1}$) from 5 to 95\% of the
  peak emission in steps of 10\%.}
\label{hco+_spectra} 
\end{figure*}

\subsection{Core masses and column densities}
\label{continuum}

The masses and column densities of the compact structures within subregions A to F can be estimated based on the 3.6\,mm continuum
emission (Fig.~\ref{mom0}, top left panel). Assuming optically thin
dust continuum emission at mm wavelengths, we followed the approach
originally outlined in \citet{hildebrand1983} with the equations for
gas mass and column density presented in \citet{schuller2009} as

\begin{eqnarray}
M_{\rm{gas}} &=& \frac{d^2S_{\rm{int}}R}{B(T_d)\kappa_{\nu}}\\
N_{\rm{gas}} &=& \frac{S_{\rm{peak}}R}{B(T_d)\Omega\kappa_{\nu}\mu m_{\rm{H}}}
,\end{eqnarray}

with the distance $d$ (1.4\,kpc; \citealt{rygl2012}), the integrated
and peak flux densities $S_{\rm{int}}$ and $S_{\rm{peak}}$, the
gas-to-dust mass ratio $R$ (150; \citealt{draine2011}), the Planck
function $B(T_d)$ depending on the dust temperature $T_d$, the dust
absorption coefficient $\kappa_{\nu}$, the beam size $\Omega$, the
mean molecular weight $\mu$ ($\sim$2.33; \citealt{kauffmann2008}), and
the mass of the hydrogen atom $m_{\rm{H}}$. The dust absorption
coefficient $\kappa_{\nu}$ was extrapolated to 3.6\,mm from
\citet{ossenkopf1994} for dust with thin ice mantles at densities of
$10^5$\,cm$^{-3}$. The dust temperatures $T_d$ were estimated from the
Herschel dust map (Fig.~\ref{dr20}, \citealt{marsh2017}). We used $T_d$
of 29 and 27\,K for regions A and B, respectively. As outlined in section \ref{temperatures} where the temperatures are estimated from the HCN/HNC ratio, the temperatures derived from dust emission are likely lower limits for the more evolved regions A and B. Therefore, for comparison, we estimated the parameters for these two regions also with a higher temperature of 50\,K (Table \ref{core_parameters}). For the four
remaining regions, we uniformly used 20\,K. The nominal 3$\sigma$ point
source mass and column density sensitivities for 20\,K are 0.6\,M$_{\odot}$
and $6\times 10^{22}$\,cm$^{-2}$.

Another important information at long wavelength is how much free-free
emission contributes to the measured flux densities. The GLOSTAR
Galactic plane survey observed the Cygnus X region at 5.8\,GHz with
the Very Large Array (VLA) in two different configurations. The
compact D-array observations result in the $19''$ resolution image
presented in Fig.~\ref{dr20} that shows the extended H{\sc ii} region
encompassing our mm peak positions A and B. Estimating the free-free
emission from these data in comparison to our much smaller 3.6\,mm
wavelength beam size of $3.45''\times 2.77''$ will overestimate the
free-free contribution at our wavelengths. Therefore, we also used the
high-resolution B-array GLOSTAR data that have a synthesized beam of
$1.5''$ at 5.8\,GHz. These higher-resolution data detect an almost
unresolved point source toward peak A and no compact cm emission
toward peak B. Using these data will hence underestimate the free-free
contribution to our 3.6\,mm data. To bracket the free-free
contribution, we measured the 5.8\,GHz flux for both datasets and
extrapolated these fluxes from 5.8\,GHz to 82.028\,GHz assuming
optically thin emission\footnote{At the lower frequencies close to
  5.8\,GHz, the optically thin assumption may not even be
  valid. Hence, the extrapolated free-free contributions could also be
  lower limits.} via $S_{\rm{free-free}}(82\rm{GHz})\sim
S_{\rm{free-free}}(5.8\rm{GHz})\times\left(\frac{82}{5.8}\right)^{-0.1}$. The
original integrated and peak flux densities $S_{\rm{int}}$ and
$S_{\rm{peak}}$ as well as the corrected values corresponding to the
D- and B-array VLA observations are listed in Table
\ref{core_parameters}.

Based on these assumptions and input information, the derived gas column
densities $N_{\rm{gas}}$ and gas masses $M_{\rm{gas}}$ are presented
in Table \ref{core_parameters}. The masses are derived within the
6$\sigma$ contours around regions A to F (Fig.~\ref{mom0}). For cores A and B, ranges
are given based on the different free-free subtraction approaches and temperatures
outlined above. The regions have very high peak column densities
between a few times $10^{23}$ and a few times
$10^{24}$\,cm$^{-2}$. The measured masses of the six main cores range
between 3.8\,M$_{\odot}$ and a potential maximum around
200\,M$_{\odot}$ (depending on the free-free contribution and temperature, in
particular, for core A). Comparing the sum of the masses shown in Table
\ref{core_parameters} with the entire mass of the region of
$\sim$2500\,M$_{\odot}$ estimated above, we see
that for the continuum data a large fraction of the flux is filtered
out, since we have no continuum short spacing information in our
data. Depending on the lower or upper mass ranges for cores A and B
(Table \ref{core_parameters}), we recover roughly between 10 and 15\%
of the entire mass in our NOEMA 3.6\,mm continuum data.

\subsection{Gas kinematics}
\label{kinematics}

Figure \ref{mom1} now shows the first-moment maps (intensity-weighted
peak velocities) of a few specific molecules. We selected three
molecules that trace the more diffuse gas (HCO$^+$, H$_2$CO, and CCH)
and the dense gas (N$_2$H$^+$). Interestingly, the dense
gas tracer N$_2$H$^+$ barely shows any significant velocity gradient
of the large-scale DR20 region, but is mainly around
$-2$\,km\,s$^{-1}$. In contrast to this, the kinematic signatures of the
other shown molecules that also trace more extended and diffuse
emission are very different. While there are also large regions that
have velocities around $-2$\,km\,s$^{-1}$, we also find large areas with more
blueshifted (up to $-6$\,km\,s$^{-1}$) and more redshifted (up
to +2\,km\,s$^{-1}$) gas.

\begin{figure*}[htb]
\includegraphics[width=0.99\textwidth]{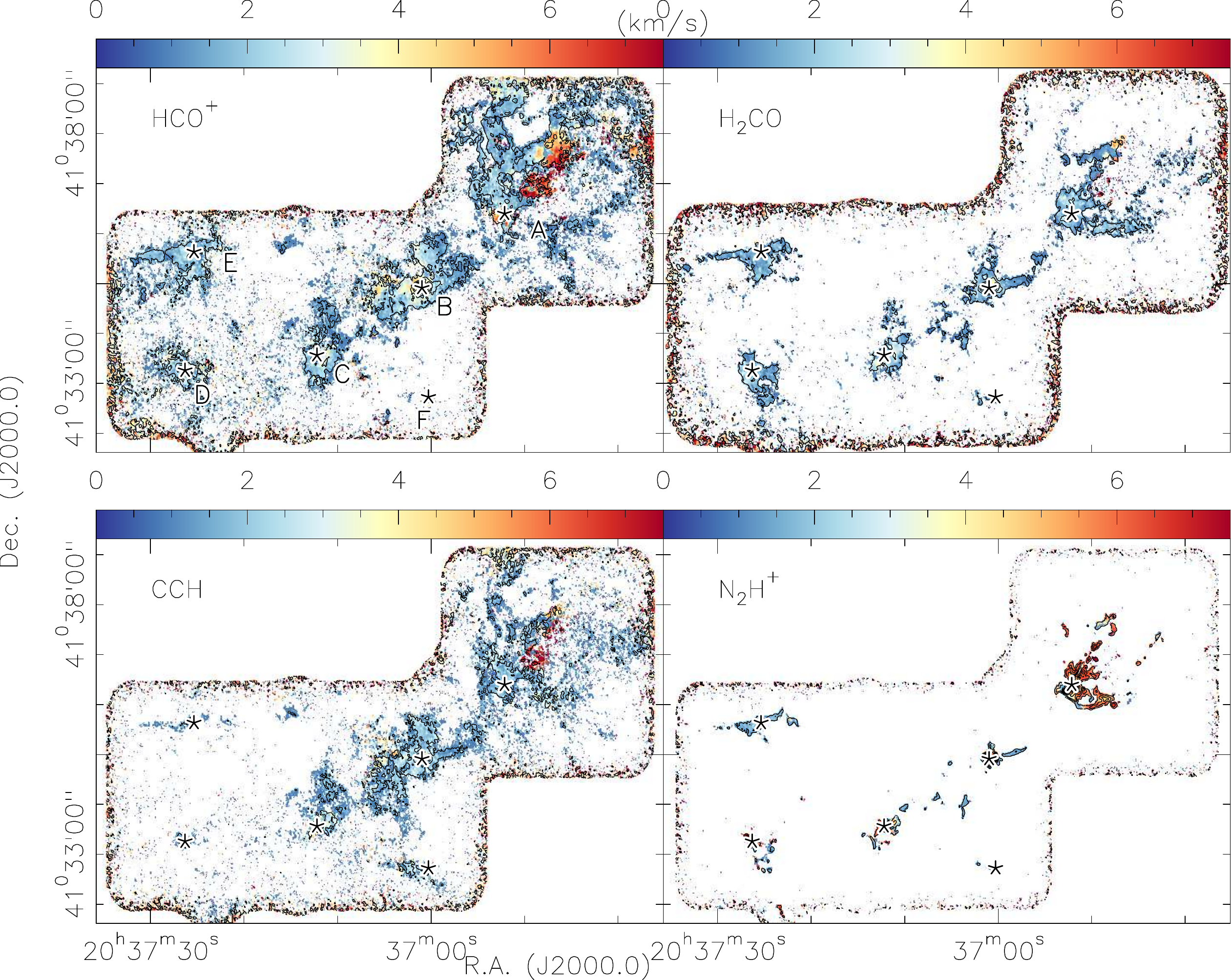}
\caption{Second-moment maps: The color scale always shows the second-moment for the labeled species. The black contours
  outline the integrated emission of the same species, always contoured  from 0.1 to 2.9\,Jy\,beam$^{-1}$\,km\,s$^{-1}$ in steps of
  0.3\,Jy\,beam$^{-1}$\,km\,s$^{-1}$. The 3.6\,mm continuum peak positions are marked with five-point stars.}
\label{mom2} 
\end{figure*}

A different way to visualize these differences is a channel
map. For this purpose, we selected the already shown HCO$^+$ and its
optically thin H$^{13}$CO$^+$ counterpart shown in Figure
\ref{hco+_channel}. We focus here on the northwestern tile around main clump A that is also the most active region with an extended
H{\sc ii} region (Figs.~\ref{dr20} and \ref{dr20_spitzer}). The
emission in the dense optically thin line H$^{13}$CO$^+$ is
different from that of the main isotopolog HCO$^+$. More precisely,
the H$^{13}$CO$^+$ line mainly emits in a small velocity regime around
$-2$\,km\,s$^{-1}$ (Fig.~\ref{hco+_channel}), similar to what is seen in the first-moment map of
N$_2$H$^+$ mentioned above. In comparison to this, the main
isotopolog HCO$^+$ is rather weak in these channels, especially west of the core A peak position (magenta five-point stars in Fig.~\ref{hco+_channel}). While this might be self-absorption due to high-optical depth, this seems unlikely because the peak-intensities at these positions west of the core A in the optically thin H$^{13}$CO$^+$ line are comparably weaker than at some other positions where the HCO$^+$ and H$^{13}$CO$^+$ emissions agree spatially and spectrally much better (Figs.~\ref{hco+_channel} and \ref{hco+_spectra}). Nevertheless, some self-absorption may still affect the HCO$^+$ emission, in particular at the highest column densities around the peak position of core A.

In contrast to H$^{13}$CO$^+$, the HCO$^+$ emission shows more 
extended as well as filamentary structures at more blue-
and more redshifted velocity channels. In particular, the blueshifted
emission between $-6$ and $-4.4$\,km\,s$^{-1}$ exhibits extended
filamentary structures that are directly connected to the main central mm and
cm continuum source. In comparison, in the redshifted
channels between 1.2 and 2.8\,km\,s$^{-1}$ , we also identify extended
and filamentary structures. These are slightly less clearly connected
to the main mm and cm source in the region, but the higher the
velocities, the closer the gas emission peaks to the central
main source.

The spectra observed in the different
parts of the region can also be studied directly. Figure \ref{hco+_spectra} shows example spectra
in the HCO$^+$(1--0) and H$^{13}$CO$^+$(1--0) lines toward selected
positions again in the northwestern part of DR20 around clump
A. Similar to the channel maps in Fig.~\ref{hco+_channel}, H$^{13}$CO$^+$ is only detected toward far fewer positions in the spectra
than the main isotopolog. In the classical picture in which
infall is diagnosed by asymmetric blue-skewed spectra of the optically thick
species (here HCO$^+$) versus centrally peaked spectra of the
optically thin counterpart (here H$^{13}$CO$^+$; e.g.,
\citealt{myers1996}), the H$^{13}$CO$^+$ line is expected to peak
in between the blue- and redshifted HCO$^+$ peaks. However, this
signature is observed nowhere in the spectra of Figure
\ref{hco+_spectra}. Therefore, the data of that region around clump A
do not indicate classical global infall motions. The two velocities
rather represent two different gas components in that region.

Because the blue- and redshifted molecular gas is 
connected via
extended filamentary structures with the central star-forming core, these structures might be filamentary feeding
channels in which gas might be accreted onto the central star-forming
complex. Alternatively, these filaments may even be feedback signatures
from the central source. We try to distinguish these two sources below.

\begin{figure*}[htb]
\includegraphics[width=0.99\textwidth]{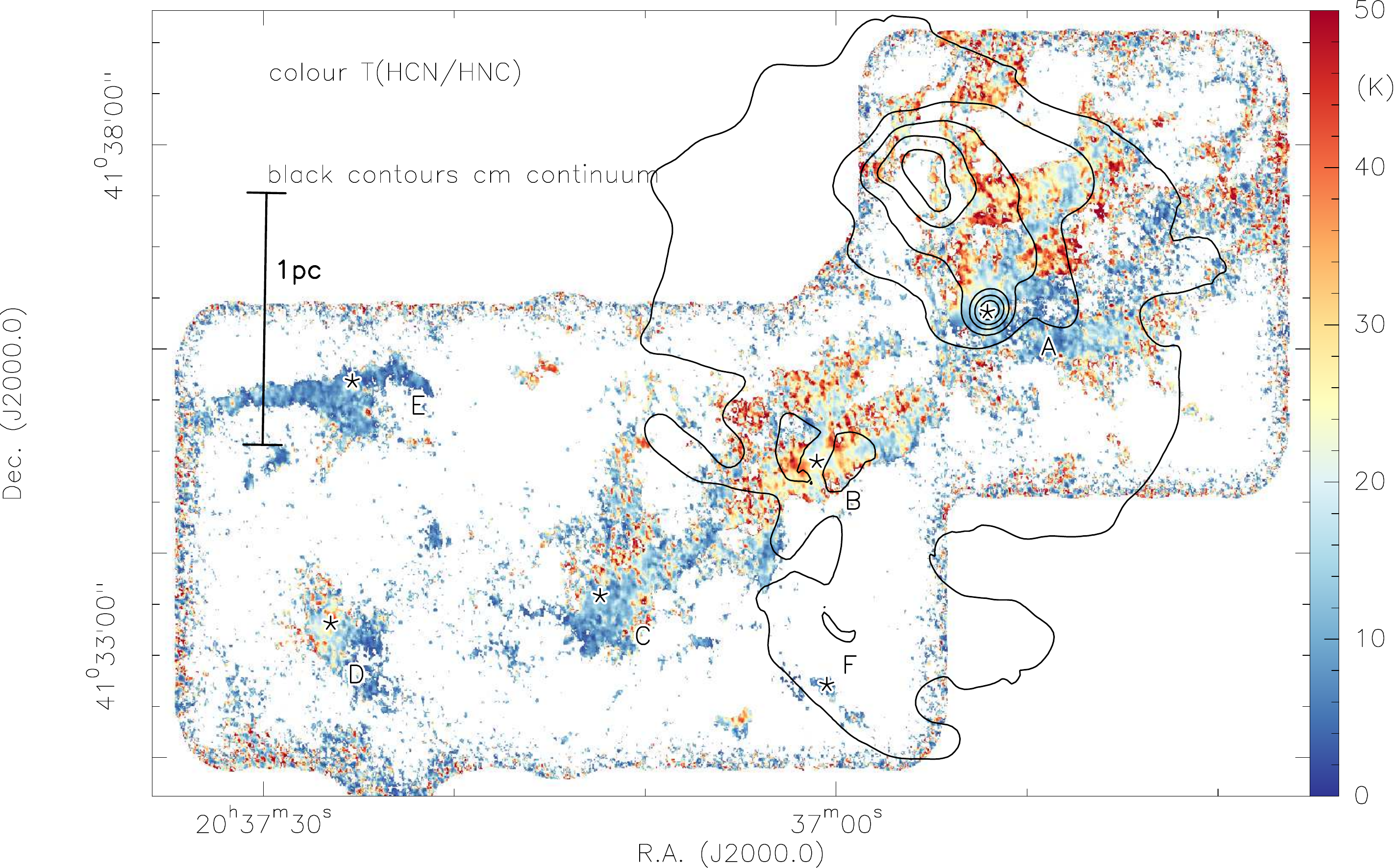}
\caption{Temperature map from the HCN/HNC ratio. The color-scale shows
  the temperature map derived from the integrated intensity ratios of
  the HCN(1--0) and HNC(1--0) line following \citet{hacar2020}. The
  black contours show the GLOSTAR cm continuum emission from 20 to
  100\,mJy\,beam$^{-1}$ in steps of 10\,mJy\,beam$^{-1}$. The 3.6\,mm
  continuum positions are marked as five-point stars. A linear
  scale-bar is shown at the left, and the six main regions A to F are
  labeled.}
\label{temp} 
\end{figure*} 

\begin{figure}[htb]
\includegraphics[width=0.49\textwidth]{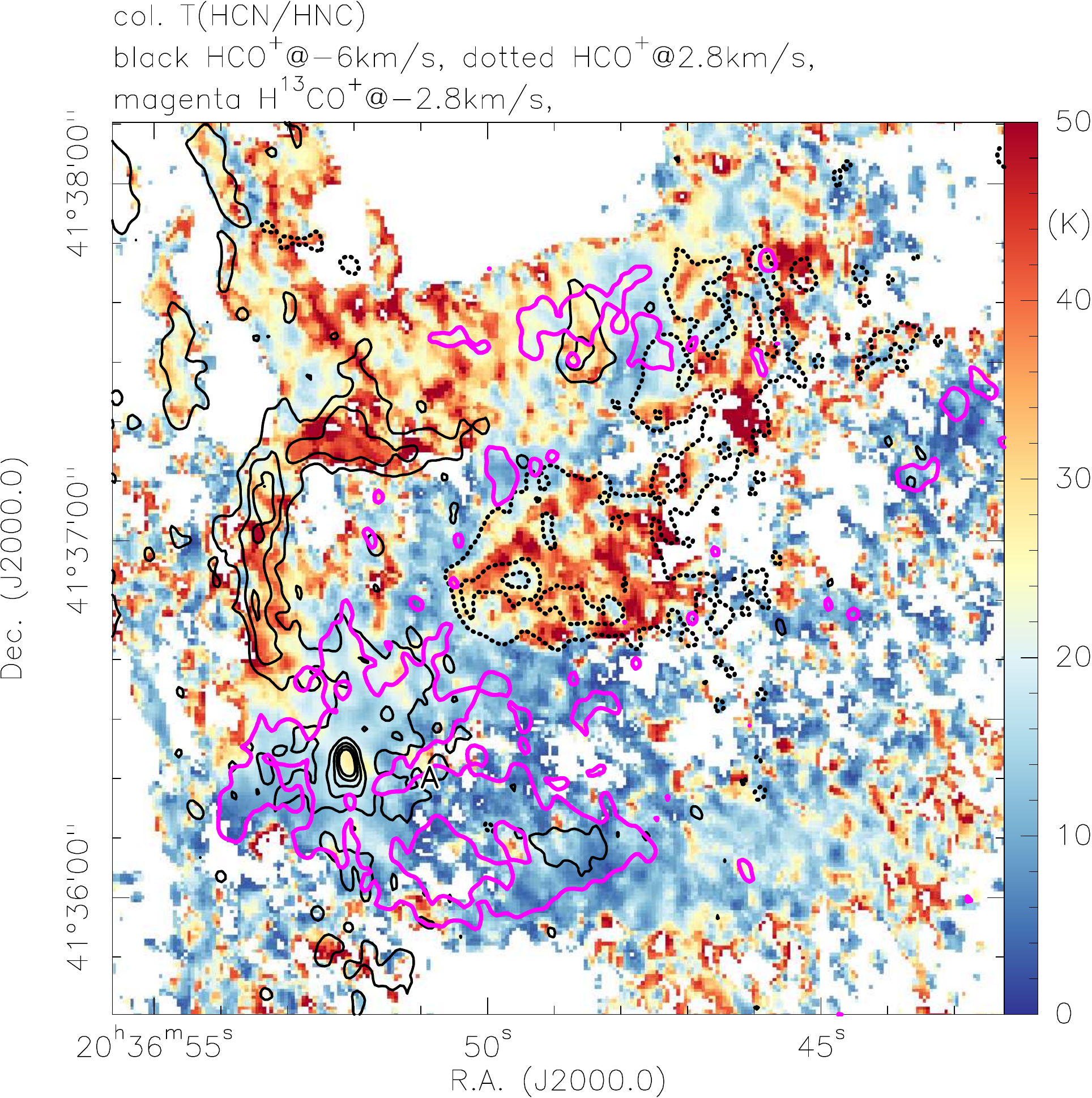}
\caption{Zoom of the temperature map for the northwestern area around clump A. The color-scale again shows the temperature map derived from the
  integrated intensity ratios of the HCN(1--0) and HNC(1--0) line
  following \citet{hacar2020}. The contours outline HCO$^+$ and H$^{13}$CO$^+$ emission structures at specific velocities (Fig.~\ref{hco+_channel}). Full black contours plot HCO$^+$at $-6$km\,s$^{-1}$, dotted contours show HCO$^+$at 2.8km\,s$^{-1}$, and magenta contours show H$^{13}$CO$^+$at $-2.8$km\,s$^{-1}$. Contour levels for HCO$^+$ start at 5$\sigma$ and continue in 10$\sigma$ steps. For H$^{13}$CO$^+$ only the 5$\sigma$ contour of the 6\,mJy\,beam$^{-1}$ is shown.}
\label{temp_zoom} 
\end{figure} 

The extended ionized gas as traced by the GLOSTAR cm continuum
emission (yellow contours in Fig.~\ref{hco+_channel}) does exhibit
emission in the northern direction that might be associated
with the blueshifted gas. However, on the redshifted side,
associations between the cm emission and the ionized gas appear less
likely. As an additional indicator for the nature of the gas flow,  the velocity dispersion may be studied, which is shown as second-moment maps
(intensity-weighted velocity dispersion) in Figure \ref{mom2}. The velocity dispersion is largely uniformly low around
$2-3$\,km\,s$^{-1}$ in the northern filamentary structures at very
negative velocities as well (Fig.~\ref{hco+_channel}). The only area with
a broader velocity dispersion is the region northwest of clump A,
where we had previously identified the two velocity components. Hence,
the higher values in the second-moment map there do not indicate broader
line widths; this is just mimicked by the two velocity
components. In the framework of the above question whether the two
velocity components may be caused by infall or by feedback, feedback
would in general cause shocks and broader lines. Therefore, the on
average narrow line widths are indicative that the two velocity
components may indeed rather belong to pristine gas streamers that may
be associated with the cloud and star formation processes. However, as we show in the following section \ref{temperatures}, the high-velocity gas components are spatially associated with enhanced temperatures. Hence, expansion and feedback from the H{\sc ii} region into the pristine environmental gas may be a viable scenario to explain the high-velocity gas.

For comparison, Figure \ref{rrl} also presents the first- and second-moment
maps of the H$44\alpha$ emission toward the strongest continuum
source A. Because of the lower sensitivity to the recombination line
emission at mm wavelengths, the line is only detected over an extent
of a few beams around the main source A. While the second-moment map
typically shows broad almost thermal velocity dispersion values
around 20\,km\,s$^{-1}$, the first-moment map in the middle panel of
Figure \ref{rrl} depicts a velocity gradient in the northwest-southeast
direction. In comparison, we do not identify a clear bipolar molecular outflow structure in the SiO emission. Interestingly, the blueshifted emission is toward the
northwest, opposite to the redshifted molecular filaments northeast
of region A that were discussed in the previous paragraphs. 

\section{Discussion}
\label{discussion}

\subsection{Gas temperatures}
\label{temperatures}

The temperatures of the gas provide important physical information for a general understanding of the region. While the Herschel data provide
us with a dust temperature map at $12''$ resolution (Fig.~\ref{dr20};
\citealt{marsh2017}), we would like to derive a temperature map at
higher angular resolution from the gas lines that we have observed. 
Although our setup comprises a few molecules with
several lines that may be used as temperature indicator, for example, from
CH$_3$CN or CH$_3$OH, emission of these molecules is typically not
that widespread, but is more confined to the central densest regions. This
is also the case in DR20; see Figure \ref{mom0_3}. However,
\citet{hacar2020} recently suggested that the ratio of the HCN(1--0) and HNC(1--0)
lines might be a good temperature measure, in particular in the regime
between 15 and 40\,K. Using the integrated intensity images shown in
Fig.~\ref{mom0_2}, we created a ratio map of the two and converted
this into temperatures following the recipe outlined in
\citet{hacar2020},

\begin{eqnarray}
T = 10 \times \left[\frac{I(\rm{HCN})}{I(\rm{HNC})}\right] \hspace{0.3cm} {\rm{if}}  \hspace{0.3cm} \frac{I(\rm{HCN})}{I(\rm{HNC})} \leq 4 \\
  T = 3 \times \left[\frac{I(\rm{HCN})}{I(\rm{HNC})}-4\right]+40 \hspace{0.3cm} {\rm{if}}  \hspace{0.3cm} \frac{I(\rm{HCN})}{I(\rm{HNC})} > 4 
.\end{eqnarray}

The resulting temperature map is presented in
Fig.~\ref{temp}. \citet{hacar2020} estimated the uncertainties of the
estimated temperatures for $10<T<40$\,K to be $\Delta T\approx 5$\,K, and
above $T>40$\,K to be $\Delta T\approx 10$\,K. A few general features can
be pointed out in the DR20 region: In the more quiescent regions C to
F, the gas temperatures are typically low, about 20\,K or even
lower. In contrast to this, in regions of strong cm continuum emission, as
outlined by the GLOSTAR contours in Fig.~\ref{temp}, the temperatures
rise to values above 40\,K. This is particularly prominent north of
region A, as well as around region B. The
GLOSTAR cm continuum and Spitzer mid-infrared images in
Figs.~\ref{dr20} and \ref{dr20_spitzer} conversely show that the regions of elevated
temperature in the HCN/HNC temperature spatially coincide with regions of
strong cm continuum and mid-infrared emission, which means ongoing or even
past star formation events. This star formation can excite the
environmental gas and hence explain the higher temperatures in the
HCN/HNC temperature maps in these regions. 

In addition to this, when the temperature map is compared to the
HCO$^+$ channel maps (Fig.~\ref{hco+_channel}), the warm regions north
of clump A appear to be associated with specific velocity
components. To examine this in more detail, Fig.~\ref{temp_zoom}
presents a zoom of the temperature map toward the north-western part
of DR20 around clump A. We overplot the contours of specific HCO$^+$
and H$^{13}$CO$^+$ channel and find clear associations, especially of
elevated temperatures with high-velocity HCO$^+$ gas. The full black
contours in Fig.~\ref{temp_zoom} show that the blueshifted filamentary
HCO$^+$ emission at $-6$\,km\,s$^{-1}$ is clearly associated with the
warm gas north of clump A. Similarly, the redshifted gas at
+2.8\,km\,s$^{-1}$ is spatially correlated with the warm gas northwest
of clump A (dotted contours in Fig.~\ref{temp_zoom}). In contrast to
this, the main dense gas structure visible in the H$^{13}$CO$^+$
emission at $-2.8$\,km\,s$^{-1}$ is spatially correlated with the cold
gas (magenta contours in Fig.~\ref{temp_zoom}). Since clump A drives
an expanding H{\sc ii} region that is visible in the cm continuum data
(e.g., yellow contours in Fig.~\ref{hco+_channel}), a plausible
interpretation for the spatial correlation between high temperatures
and high-velocity gas could be that the expanding H{\sc ii} region
heats up the gas and at the same time pushes the remnant structures of
the envelope to higher velocities. In contrast to this, the
higher-density tracer H$^{13}$CO$^+$ may trace the parts of the
envelope that are still less affected by the expanding H{\sc ii}
region.

\begin{figure}[htb]
\includegraphics[width=0.49\textwidth]{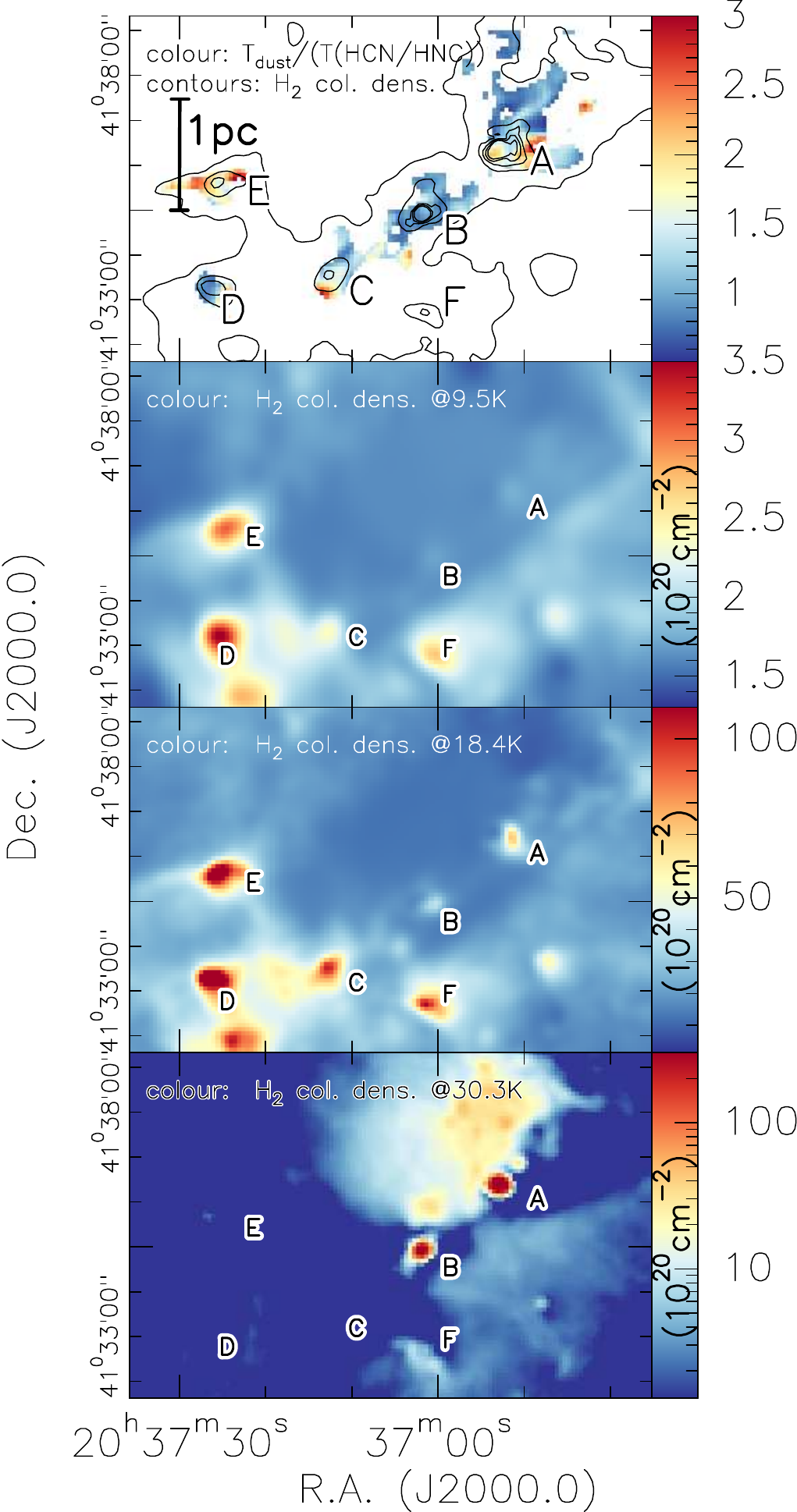}
\caption{Temperature ratio map and column density maps at different
  temperatures. {\it Top panel:} Ratio map of the temperatures derived
  from the Herschel far-infrared data (Fig.~\ref{dr20};
  \citealt{marsh2017}) vs the temperature map obtained from the
  HCN/HNC ratio (Fig.~\ref{temp}; top panel). For this ratio map, the
  HCN and HNC data were first smoothed to the $12''$ resolution of the
  dust temperature map. The black contours show the H$_2$ column
  density map derived by \citet{marsh2017} at contour levels between
  $10^{22}$ and $5\times 10^{22}$\,cm$^{-2}$ (step $1\times
  10^{22}$\,cm$^{-2}$). A linear scale-bar is shown to the left. The
  {\it bottom three panels} show the H$_2$ column densities derived at
  different temperatures (as marked in the panels) as outlined in
  \citet{marsh2017}. The six main regions A to F are marked in all
  panels.}
\label{temp_ratio} 
\end{figure} 

Toward the peak positions of regions A and B, we find
elevated temperatures, as expected, but especially toward B, they are slightly below the
environmental higher temperatures. This is partly in contrast to the
temperature map derived from the far-infrared continuum data
(Fig.~\ref{dr20}; \citealt{marsh2017}) where the highest temperatures
are found toward the peak positions A and B. We point out that
although elevated above 20\,K, the absolute values of the Herschel
temperature map even in the warm regions north of region A are lower
than the HCN/HNC derived values. This can be explained by the combined
effect of the lower angular resolution of the Herschel data and
the fact that the Herschel far-infrared wavelengths between 70 and
500\,$\mu$m are not a good tracer of gas warmer than
40\,K. Nevertheless, the gradients and temperature trends in the
Herschel map reflect the large-scale temperature structure of the
region well. Therefore, the estimated lower temperatures toward the A
and B peak positions in the HCN/HNC derived temperature compared to
the surroundings of A and B are most likely an artifact because
toward these positions, the H$_2$ column densities rise above values of
$10^{23}$\,cm$^{-2}$ (section \ref{continuum}), and the HCN and HNC
emission becomes optically thick. In this regime, the ratio of
the two lines decreases again and saturates at values corresponding to
lower temperatures. \citet{hacar2020} argued that up to extinctions
$A_v\approx 100$, corresponding to H$_2$ column densities of $\sim
1\times 10^{23}$\,cm$^{-2}$, the HCN/HNC ratio could still be a
reasonable column density tracer. However, toward the peak positions
of A and B, we are even above that threshold (Table
\ref{core_parameters}). We return to these high-column density
regions below by investigating the H$^{13}$CN/HN$^{13}$C ratio.

For a direct comparison of the temperature map derived from
the Herschel far-infrared data (Fig.~\ref{dr20}, \citealt{marsh2017})
and our higher-resolution HCN/HNC temperature map, we smoothed the
HCN/HNC temperature map to the spatial resolution of $12''$ of the
dust temperature map and placed them on the same grid. The ratio of these
two maps is presented in Fig.~\ref{temp_ratio} (top panel). It is
encouraging to see that over large areas of the map, the ratio of
dust temperature and HCN/HNC temperature is close to 1. This is
particularly clear north of region A, in almost the entire
region B, and in large parts of C and D. Excluding some edges around C and
D, the only exceptions of a ratio on the order of unity are the peak
of region A, some area slightly west of A, and almost the entire
region E. While the difference toward the peak of A can most likely
be explained by the high column densities toward that position and
the accordingly increased optical depth in the HCN and HNC lines, this seems
less likely to be the case for the high ratios west of A and in E.

\begin{figure}[htb]
\includegraphics[width=0.49\textwidth]{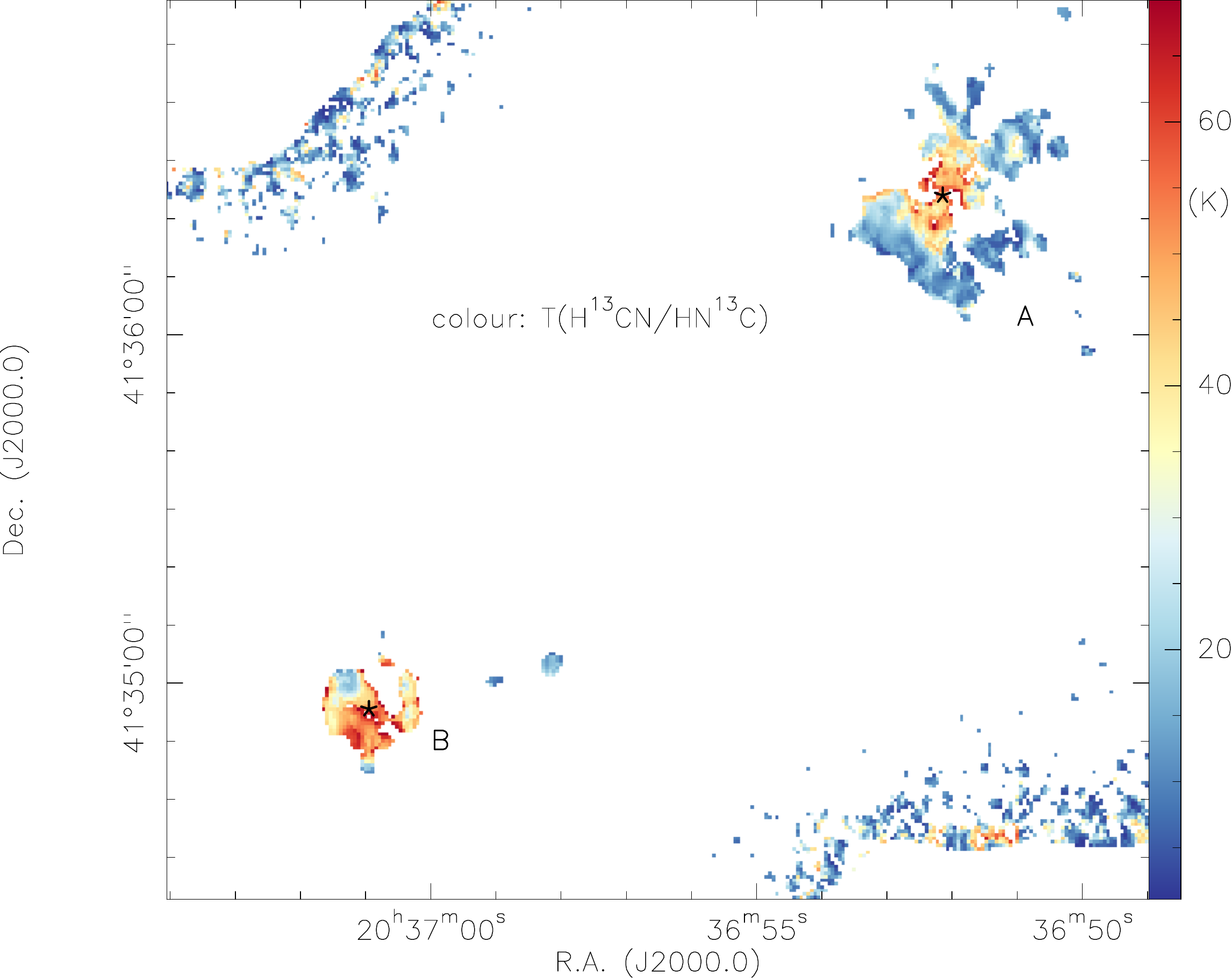}
\includegraphics[width=0.49\textwidth]{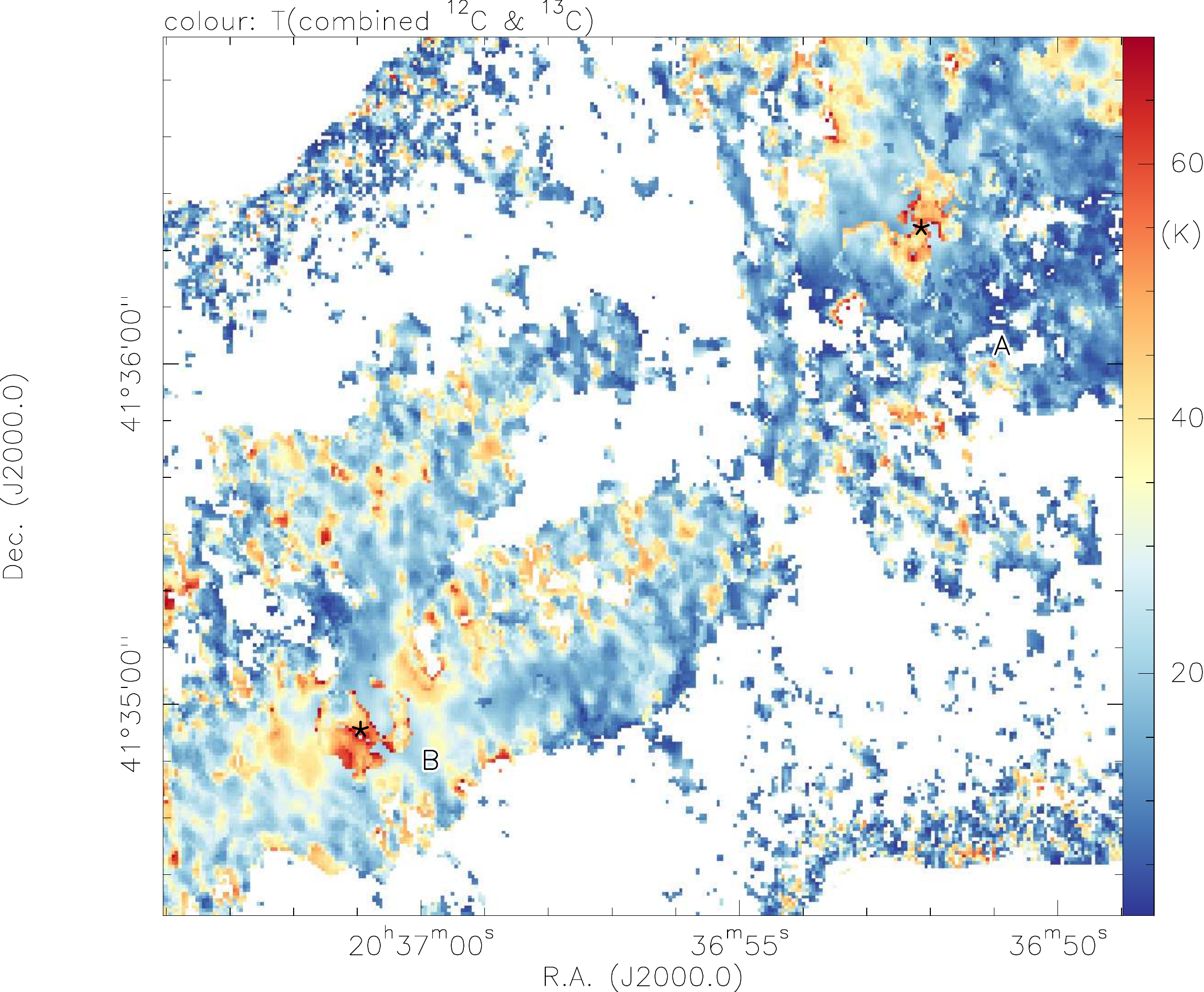}
\caption{Temperature maps using also the H$^{13}$CN/HN$^{13}$C
  ratio. {\it Top panel:} Temperature map derived from the integrated
  intensity ratios of the H$^{13}$CN(1--0) and HN$^{13}$C(1--0) line
  following the procedure outlined in \citet{hacar2020} for the main
  isotopologs. The five-point stars mark the 3.6\,mm continuum peak
  positions of region A and B. The {\it bottom panel} presents the
  corresponding temperature map where the temperatures derived from
  HCN/HNC and H$^{13}$CN/HN$^{13}$C are combined (see main text for
  procedure).}
\label{temp2} 
\end{figure}

Qualitatively speaking the dust temperature and
HCN/HNC temperature maps in region E largely agree in the sense that low
temperatures are found in both of them. Nevertheless, quantitatively,
the estimates differ in the dust temperature map toward E, the
temperatures drop below 20\,K but never lower than 17\,K. In contrast, the HCN/HNC temperature map there exhibits largely
temperatures around 10\,K or even below, implied by ratios of about
2. A similar situation can be found east of region A, where the dust
temperature map ranges around 22\,K, and the HCN/HNC temperature map
again shows values in the 10\,K regime. The higher-than-1 ratio between the
dust and HCN/HNC temperatures west of peak A can be explained by the
radiation in the H{\sc ii} region. This radiation may excite 
outer
layers of the dust environment, and by this, increase the average
line-of-sight dust temperature. In contrast, these warmer
layers may not be associated with higher-density gas traced by
HCN/HNC. These molecular lines rather trace the denser and cooler gas,
explaining the higher ratios seen west of A in Fig.~\ref{temp_ratio}.

The PPMAP approach (point process map) by \citet{marsh2015,marsh2017} for estimating column densities
and temperatures from Herschel data has  not only been applied to the
integrated H$_2$ column density map and the line-of-sight averaged
temperature maps, but also provides H$_2$ column density maps for
different temperatures (12 bins between 8 and 50\,K). The three bottom
panels of Fig.~\ref{temp_ratio} now show these H$_2$ column densities
at three selected temperatures of 9.5, 18.4, and 30.3\,K. The differences
between these maps is large in that sources E and F are
mainly found to have  very low temperatures, whereas A and B have
high temperatures. In source E, the dense central
region receives strong contributions of cold gas and dust. However, in the
line-of-sight averaged temperature map (Fig.~\ref{dr20}), the outer warmer layers of the region
contribute so strongly that temperatures below 17\,K cannot be 
attained. 
In contrast to this, HCN and HNC have critical densities
above $10^5$\,cm$^{-3}$ and even effective excitation densities above
$10^3$\,cm$^{-3}$ \citep{shirley2015}. Hence, these lines
preferentially trace the higher density, and in region E, also colder gas. Therefore, measuring lower temperatures via the
HCN/HNC ratio is plausible.

We return to the potential high optical depth, especially toward the
peak positions A and B. We can also explore the emission from the rarer isotopologs
H$^{13}$CN and HN$^{13}$C. As shown in Fig.~\ref{mom0_2}, these rarer
isotopologs are mainly detected toward the two main peak regions A
and B. This is sufficient, however, because these regions also have the
highest optical depth, where the main isotopologs discussed before
are most prone to saturation and hence deliver temperature
estimates with high uncertainties. We now applied exactly the same procedure to the integrated
intensity ratio maps of H$^{13}$CN and HN$^{13}$C and derived a
new temperature map in particular towards regions A and B. The
resulting temperature distribution is shown in Fig.~\ref{temp2} (top
panel). As expected, the highest temperatures are now found
toward the peak positions of A, where the most luminous embedded
sources should be located. Hence qualitatively speaking, using the
rarer $^{13}$C isotopologs appears to be a reasonable option to
estimate temperatures toward high column density regions as well.

\begin{figure*}[htb]
\includegraphics[width=0.99\textwidth]{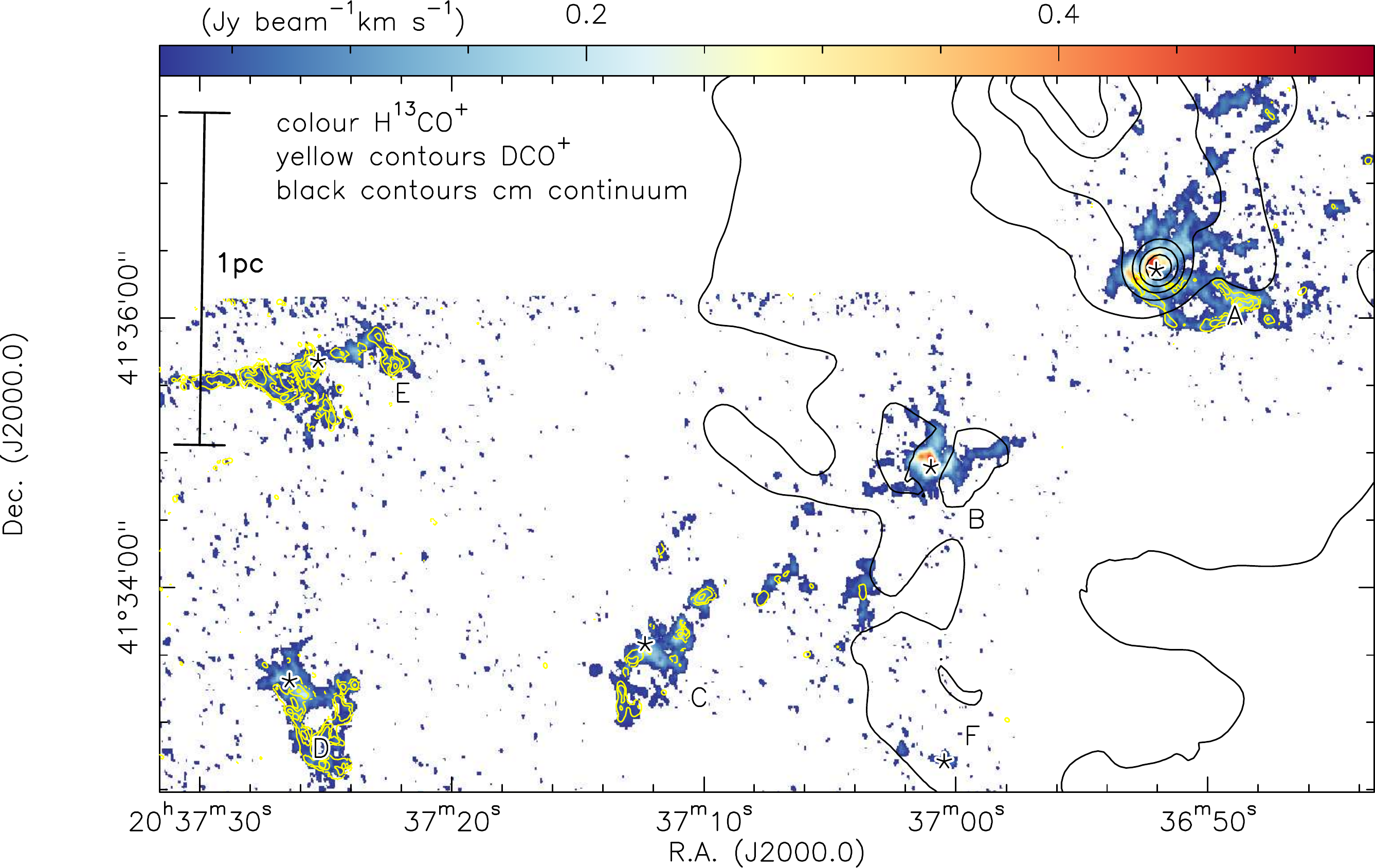}
\caption{Integrated DCO$^+(1-0)$ contours in yellow on
  H$^{13}$CO$^+(1-0)$ emission in color. The integration regimes for
  the two molecular transitions are from -7 to 3\,km\,s$^{-1}$. The
  yellow DCO$^+$ contours are from 50 to
  500\,mJy\,beam$^{-1}$km\,s$^{-1}$ in steps of
  50\,mJy\,beam$^{-1}$km\,s$^{-1}$. The black contours show the
  GLOSTAR cm continuum emission from 20 to 100\,mJy\,beam$^{-1}$ in
  steps of 10\,mJy\,beam$^{-1}$. A linear scale-bar is shown to the
  left, and the six main regions A to F are labeled and marked by
  five-point stars.}
\label{dco+} 
\end{figure*}  

As a final test, we combined the results obtained with the main
isotopolog lines HCN/HNC with those from their rarer $^{13}$C
replacements H$^{13}$CN/HN$^{13}$C. Assuming that the chemical
behavior of the $^{12}$C and the $^{13}$C isotopologs is similar, we
used the H$^{13}$CN/HN$^{13}$C ratio where detectable and the HCN/HNC ratio
for the remaining parts of the map. The result of this brute-force
combination is presented in Fig.~\ref{temp2} (bottom
panel). Interestingly, the transition between the two temperature
derivations is barely recognizable in this combined map. This can be
considered as an indicator that deriving the temperature distribution
by this combined approach can be reasonable. While the
H$^{13}$CN/HN$^{13}$C ratio can be used for the highest column density
regions, the HCN/HNC ratio can be employed for the remaining ga,s as
introduced by \citet{hacar2020}.

We are aware that we do not present  an in-depth analysis of the
chemical background of this potential thermometer as well. Our aim was
to further explore the possibilities of using the HCN/HNC and
H$^{13}$CN/HN$^{13}$C ratios as gas thermometer in the 3\,mm
wavelength band.

\subsection{Deuteration}
\label{deuteration}

With our spectral setup, we also cover several lines from deuterium substitute isotopologs, in particular from  DCO$^+$, DCN, DNC, and NH$_2$D
(Table \ref{rms}). The integrated emission maps of these species
presented in Figures \ref{mom0}, \ref{mom0_2}, and \ref{mom0_3} show
that in the
DR20 region, the DCO$^+(1-0)$ emission is strongest and most extended. To highlight some of the questions that can be addressed
with deuterated molecules, we therefore focus on the DCO$^+(1-0)$
emission. It has also been shown in the framework of the IRAM
CORE large program that DCO$^+$ is an excellent tracer of the dynamics
of the gas in very young star-forming regions
\citep{beuther2021,gieser2022}. Because the DR20 region contains sources in  several
evolutionary stages, it can provide the ideal laboratory for testing whether DCO$^+$ indeed
only traces early evolutionary stages or if more evolved regions
can also be reasonably well studied by this species.

Figure \ref{dco+} shows an overlay of the integrated DCO$^+$ emission
in comparison with its non-deuterated counterpart H$^{13}$CO$^+$ (which
we used 
instead of HCO$^+$ because of its lower
abundance and hence 
lower optical depth). While there is a
clear correlation between the distributions of the deuterated and nondeuterated species
for the main cores C, D, and E, the correlation is less clear for the
two more evolved regions A and B (F is too weak to be clearly detected
in any of the two tracers). While DCO$^+$ remains entirely undetected
toward region B, toward region A, it is at least detected at the
southern boundary of the region, offset from the main 3.6\,mm
continuum emission. It is interesting to note that also towards
regions C and D, the DCO$^+$ emission is found to be offset from the main
3.6\,mm continuum peak positions. Only toward region E are
H$^{13}$CO$^+$, DCO$^+$ and 3.6\,mm continuum emission relatively well correlated spatially (Fig.~\ref{dco+}). While the
anticorrelation of DCO$^+$ with most of the continuum peaks is likely
associated with high temperatures in these regions (see below), the
nondetection of extended 3.6\,mm continuum emission in regions of
strong DCO$^+$ emission may also be related to the fact that in
contrast to the spectral line data, where we combined NOEMA and 30\,m
data, we lack the complementary short
spacing data  for the 3.6 mm continuum and hence cannot recover extended emission.

One physical way to interpret these results is based on a comparison
with the gas and dust temperature maps presented in Figures \ref{temp}
and \ref{dr20}.  Regions C, D, and E clearly 
have the
lowest temperatures in the two temperature maps. Even
the deuterated emission in the vicinity of peak A lies south and
southwest of the strong mm and cm continuum emission where the gas
temperatures are below 20\,K. Hence, a temperature regime below 20\,K
clearly seems conducive for detecting deuterated DCO$^+$
emission. This is in line with other DCO$^+$ observations and
modeling, for example, in the Horsehead nebula \citep{pety2007} or even in
the Orion bar, where the deuteration in DCO$^+$ is comparably low at
the higher temperatures in this photon-dominated region
\citep{parise2009}. The fact that the continuum peak E is associated
with DCO$^+$ emission implies that this region is cold and has high
gas column densities.

\begin{figure}[htb]
\includegraphics[width=0.49\textwidth]{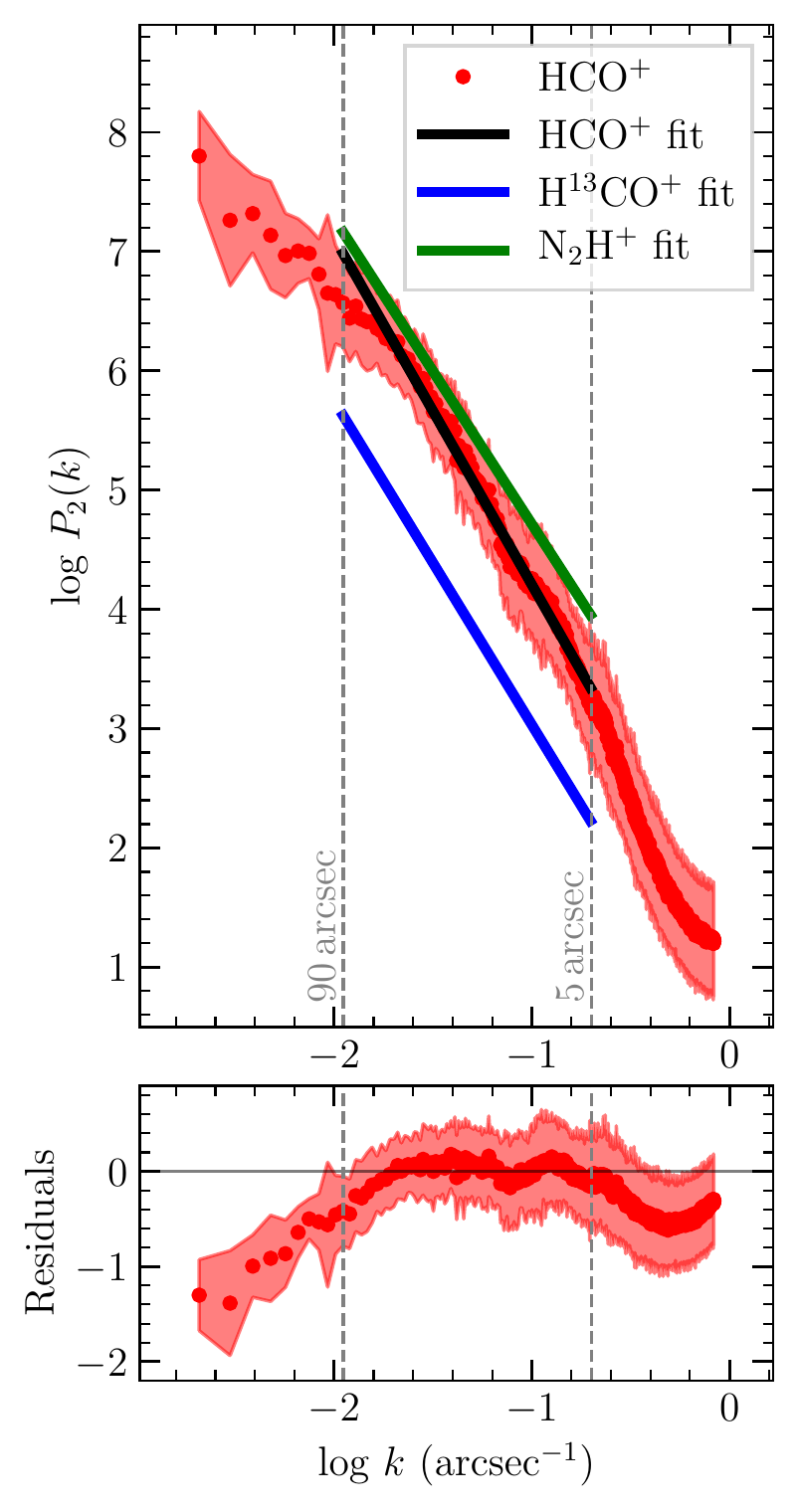}
\caption{Power spectra derived in DR20 for HCO$^+(1-0)$, H$^{13}$CO$^+(1-0)$ and N$_2$H$^+(1-0)$. The dots and colored area correspond to the HCO$^+$ data and their $1\sigma$ uncertainty. The black line is the fit to the data between $5''$ and $90''$. For comparison, we show the fits to the corresponding H$^{13}$CO$^+$ and N$_2$H$^+$ in blue and green. The bottom panel presents the residuals of the HCO$^+$ fit.}
\label{powerspectra} 
\end{figure} 

\subsection{Spatial power spectra of the intensity distributions of the dense gas tracer}

Power spectra of integrated intensity emission or entire spectral data
cubes are used to infer the turbulent and gravitational
properties of the gas in the ISM and associated with star formation
(e.g.,
\citealt{weizsaecker1951,lazarian2000,elmegreen2004,falceta-goncalves2015,burkhart2010,burkhart2021}). The
power spectrum of incompressible turbulence follows a power law of
$-11/3$ \citep{kolmogorov1941} that can be recovered by observations
of an optically thin tracer of the turbulent gas like HI (e.g.,
\citealt{lazarian2000,lazarian2002,stanimirovic2001,miville-deschenes2003,burkhart2013b}). Optical
depth effects can saturate this slope at $-3$ (e.g.,
\citealt{lazarian2004,burkhart2013b}). Other effects can flatten this
slope, and \citet{burkhart2010} showed that the slope of the power
spectrum becomes shallower for increasing Mach number because stronger
turbulence creates more density irregularities and hence more energy
on smaller scales (see also \citealt{hennebelle2012}). In addition to
this, self-gravity flattens the slopes of the power spectra even more
(e.g., \citealt{collins2012,federrath2013,burkhart2015}). This latter
picture was recently observationally confirmed by multitracer
observations and analysis of the Perseus molecular cloud:
\citet{pingel2018} investigated the power spectra of HI, $^{12}$CO,
$^{13}$CO, and dust data of the Perseus cloud and reported
power-law slopes of $-$3.2, $-$3.1,$-$ 2.9, and $-$2.7 for the four
tracers, respectively. Hence, tracers of increasing density typically
associated with increasing star formation activity exhibit
increasingly shallower power-law slopes.

This latter scenario can now be investigated further with our MIOP
observations in Cygnus X. While at our high angular resolution we lack data for the complementary low-column density tracers such as HI or CO,
we observe a broad range of higher-density gas tracers. In the
following, we concentrate on the optically thick and thin pair of
HCO$^+(1-0)$/H$^{13}$CO$^+(1-0)$ and on the N$_2$H$^+(1-0)$ line
emission. While all three lines have critical densities of about
$4\times 10^4$\,cm$^{3}$, their effective excitation densities at
20\,K span a broader range with $5.3\times 10^2$\,cm$^{3}$, $2.2\times
10^4$\,cm$^{3}$,  and $5.5\times 10^3$\,cm$^{3}$, respectively
\citep{shirley2015}. Hence, all lines trace dense gas  in which star
formation activity and correspondingly gravity are expected to dominate.

To extract the slopes of the power spectra for these three tracers for the entire DR20 region mapped here, we
used the python package {\sf TurbuStat} \citep{koch2019}. To avoid any
artifacts caused by noise, we used the images presented in Figures \ref{mom0}
and \ref{mom0_2} with the modification that we increased the threshold
for creating the integrated intensity maps to $5\sigma$, and we masked
the edges of the mosaics where the noise increases again. For HCO$^+$
and H$^{13}$CO$^+$, we used the standard integration range from $-7$
to +3\,km\,s$^{-1}$. For N$_2$H$^+$ we increased the integration range
to $-12$ to +6\,km\,s$^{-1}$ to cover the entire hyperfine
structure. We derived the power spectra over spatial scales between
$5''$ (larger than the angular resolution) and $90''$, roughly the
largest size-scale that can be identified in the N$_2$H$^+$ and
H$^{13}$CO$^+$ data (Figs.~\ref{mom0} and \ref{mom0_2}). The derived
power-law slopes are from 1D fits to the azimuthally
averaged 2D power spectra. With this approach, we obtain power-law slopes
of the power spectra for HCO$^+$, H$^{13}$CO$^+$ , and N$_2$H$^+$ of
$-2.9$, $-2.7,$ and $-2.6$, respectively (Fig.~\ref{powerspectra}). The standard deviation of the
fits to the slopes in all three cases is smaller than 0.1.

These estimates in context with the power spectra results
reported in the literature and introduced above show that the high
optical depth and comparably lower density tracer HCO$^+$ 
with $-2.9$ shows a power-law slope that is close to the saturated $-3.0$ value
discussed by \citet{lazarian2004} and \citet{burkhart2013b}. In
comparison to this, the two higher-density tracers H$^{13}$CO$^+$ and
N$_2$H$^+$ reveal consistently flatter slopes of $-2.7$ and $-2.6$
that are expected for increased Mach numbers and/or self-gravity
becoming dominant compared to the initially more turbulent gas where the
slopes are steeper (e.g., 
\citealt{burkhart2010,burkhart2015,collins2012,federrath2013}). The
slopes found here for the dense gas tracers are also comparable to
those found in other studies based on dust extinction and emission, which also trace the densest parts of the star-forming regions (e.g.,
\citealt{pingel2018} and references therein). Since the DR20 region is
an active star-forming region, flatter slopes caused by the collapse
of the gas and hence dominating gravity are very plausible.

\section{Conclusions and summary}
\label{conclusion}

We presented the outline, scope, and initial results from the large Max
Planck IRAM Observatory Program (MIOP) on star formation in Cyg
X (CASCADE). Using the the two IRAM facilities NOEMA and the 30\,m telescope,
we mapped large mosaics in the Cyg X region in the 3\,mm wavelength band
in the continuum and spectral line emission. A typical angular
resolution of $\sim 3''$ results in linear resolution elements of
$<5000$\,au. Because the data are complemented by the short-spacing
observations of the 30\,m telescope, this unique new dataset allowed us
to investigate the physical and chemical processes during star
formation from large cloud scales down to our spatial resolution
limit. Furthermore, covering a plethora of spectral lines and
continuum emission, we studied a diverse set of phenomena covering
accretion flows from large to small scales, cloud collapse,
filamentary structures, and the physical and chemical
properties of the gas associated with this prototypical large star
formation complex.

After outlining the background and scope of the program, we described
the observational setup of the project. Furthermore, we presented the
data and analysis of the first observed subarea, namely the DR20
star-forming region, to highlight first results as well as the
potential of the entire survey.

The DR20 covers a range of evolutionary stages, and the diverse set of
spectral lines allowed us to trace different physical and chemical
properties. We identified several velocity components that may
stem from accretion flows onto the most massive central star-forming
gas clump and from gas that can be affected by the evolving H{\sc ii} region. Employing the HCN/HNC intensity ratio, we derived a large-scale gas
temperature map of the region and set it into context with
temperatures derived at lower angular resolution from Herschel
far-infrared data \citep{marsh2017}. Furthermore, we explored how much
the rarer isotopolog ratio H$^{13}$CN/HN$^{13}$C may potentially be
used complementary in the highest column density parts of these
star-forming regions. An analysis of the DCO$^+$ emission showed that
its emission is almost exclusively identified in cold regions with gas
temperatures below 20\,K. In addition to this, we investigated the
slope of the power spectra for the dense gas tracers HCO$^+$,
H$^{13}$CO$^+$ , and N$_2$H$^+$. While the slope of the power spectra in
a purely turbulence-dominated regime typically observed for example in
HI emission is $-11/3$, optical depth effects can flatten it up to
$-3$. For the dense gas tracers investigated here, we found slopes
of $-2.9$, $-2.7,$ and $-2.6$, respectively. These flatter slopes are
expected for increased Mach numbers and/or self-gravity that becomes
dominant compared to the initial turbulent conditions. Because the DR20
region is actively forming stars and is hence dominated by gravity, the
flatter slopes of the power spectra agree with the physical status of the
region.

The analysis of this initial DR20 dataset outlines the enormous
potential of this MIOP on star formation in Cygnus X. CASCADE will
allow us detailed investigation of the different physical and chemical
aspects and their interrelations from the scale of the natal molecular
cloud down to the scale of accretion on the individual protostellar
cores.

\begin{acknowledgements}
The authors are grateful to the staff at the NOEMA and Pico Veleta
observatories for their support of these observations. We thank in
particular P. Chaudet, operator at the NOEMA observatory, for his
motivation and dedication in developing and testing the advanced
mosaic observing procedures employed in this project. This work is
based on observations carried out under project number L19MA with the
IRAM NOEMA Interferometer and [145-19] with the 30\,m telescope. IRAM is
supported by INSU/CNRS (France), MPG (Germany) and IGN (Spain). We
like to thank Alvaro Hacar, Blakesley Burkhart and Eric Koch for stimulating
discussions about the HCN/HNC temperature and the power spectra
fitting, respectively.  H.B.~and S.S.~acknowledge support from the
European Research Council under the Horizon 2020 Framework Program via
the ERC Consolidator Grant CSF-648505. H.B. also acknowledges support
from the Deutsche Forschungsgemeinschaft in the Collaborative Research
Center (SFB 881) “The Milky Way System” (subproject B1).
\end{acknowledgements}


\begin{thebibliography}{83}
\expandafter\ifx\csname natexlab\endcsname\relax\def\natexlab#1{#1}\fi

\bibitem[{{Albertsson} {et~al.}(2013){Albertsson}, {Semenov}, {Vasyunin},
  {Henning}, \& {Herbst}}]{albertsson2013}
{Albertsson}, T., {Semenov}, D.~A., {Vasyunin}, A.~I., {Henning}, T., \&
  {Herbst}, E. 2013, \apjs, 207, 27

\bibitem[{{Andr{\'e}} {et~al.}(2014){Andr{\'e}}, {Di Francesco},
  {Ward-Thompson}, {Inutsuka}, {Pudritz}, \& {Pineda}}]{andre2014}
{Andr{\'e}}, P., {Di Francesco}, J., {Ward-Thompson}, D., {et~al.} 2014, in
  Protostars and Planets VI, ed. H.~{Beuther}, R.~{Klessen}, C.~{Dullemond}, \&
  T.~{Henning}, 27--51

\bibitem[{{Arzoumanian} {et~al.}(2011){Arzoumanian}, {Andr{\'e}}, {Didelon},
  {K{\"o}nyves}, {Schneider}, {Men'shchikov}, {Sousbie}, {Zavagno}, {Bontemps},
  {di Francesco}, {Griffin}, {Hennemann}, {Hill}, {Kirk}, {Martin}, {Minier},
  {Molinari}, {Motte}, {Peretto}, {Pezzuto}, {Spinoglio}, {Ward-Thompson},
  {White}, \& {Wilson}}]{arzoumanian2011}
{Arzoumanian}, D., {Andr{\'e}}, P., {Didelon}, P., {et~al.} 2011, \aap, 529, L6

\bibitem[{{Beerer} {et~al.}(2010){Beerer}, {Koenig}, {Hora}, {Gutermuth},
  {Bontemps}, {Megeath}, {Schneider}, {Motte}, {Carey}, {Simon}, {Keto},
  {Smith}, {Allen}, {Fazio}, {Kraemer}, {Price}, {Mizuno}, {Adams},
  {Hern{\'a}ndez}, \& {Lucas}}]{beerer2010}
{Beerer}, I.~M., {Koenig}, X.~P., {Hora}, J.~L., {et~al.} 2010, \apj, 720, 679

\bibitem[{{Beuther} {et~al.}(2016){Beuther}, {Bihr}, {Rugel}, {Johnston},
  {Wang}, {Walter}, {Brunthaler}, {Walsh}, {Ott}, {Stil}, {Henning},
  {Schierhuber}, {Kainulainen}, {Heyer}, {Goldsmith}, {Anderson}, {Longmore},
  {Klessen}, {Glover}, {Urquhart}, {Plume}, {Ragan}, {Schneider},
  {McClure-Griffiths}, {Menten}, {Smith}, {Roy}, {Shanahan}, {Nguyen-Luong}, \&
  {Bigiel}}]{beuther2016}
{Beuther}, H., {Bihr}, S., {Rugel}, M., {et~al.} 2016, \aap, 595, A32

\bibitem[{{Beuther} {et~al.}(2021){Beuther}, {Gieser}, {Suri}, {Linz},
  {Klaassen}, {Semenov}, {Winters}, {Henning}, {Soler}, {Urquhart}, {Syed},
  {Feng}, {M{\"o}ller}, {Beltr{\'a}n}, {S{\'a}nchez-Monge}, {Longmore},
  {Peters}, {Ballesteros-Paredes}, {Schilke}, {Moscadelli}, {Palau},
  {Cesaroni}, {Lumsden}, {Pudritz}, {Wyrowski}, {Kuiper}, \&
  {Ahmadi}}]{beuther2021}
{Beuther}, H., {Gieser}, C., {Suri}, S., {et~al.} 2021, \aap, 649, A113

\bibitem[{{Beuther} {et~al.}(2020){Beuther}, {Wang}, {Soler}, {Linz},
  {Henshaw}, {Vazquez-Semadeni}, {Gomez}, {Ragan}, {Henning}, {Glover}, {Lee},
  \& {G{\"u}sten}}]{beuther2020}
{Beuther}, H., {Wang}, Y., {Soler}, J., {et~al.} 2020, \aap, 638, A44

\bibitem[{{Bonnell} \& {Bate}(2002)}]{bonnell2002}
{Bonnell}, I.~A. \& {Bate}, M.~R. 2002, \mnras, 336, 659

\bibitem[{{Bonnell} {et~al.}(2004){Bonnell}, {Vine}, \& {Bate}}]{bonnell2004}
{Bonnell}, I.~A., {Vine}, S.~G., \& {Bate}, M.~R. 2004, \mnras, 349, 735

\bibitem[{{Brunthaler} {et~al.}(2021){Brunthaler}, {Menten}, {Dzib}, {Cotton},
  {Wyrowski}, {Dokara}, {Gong}, {Medina}, {M{\"u}ller}, {Nguyen},
  {Ortiz-Le{\'o}n}, {Reich}, {Rugel}, {Urquhart}, {Winkel}, {Yang}, {Beuther},
  {Billington}, {Carrasco-Gonzalez}, {Csengeri}, {Murugeshan}, {Pandian}, \&
  {Roy}}]{brunthaler2021}
{Brunthaler}, A., {Menten}, K.~M., {Dzib}, S.~A., {et~al.} 2021, \aap, 651, A85

\bibitem[{{Burkhart}(2021)}]{burkhart2021}
{Burkhart}, B. 2021, \pasp, 133, 102001

\bibitem[{{Burkhart} {et~al.}(2015){Burkhart}, {Collins}, \&
  {Lazarian}}]{burkhart2015}
{Burkhart}, B., {Collins}, D.~C., \& {Lazarian}, A. 2015, \apj, 808, 48

\bibitem[{{Burkhart} {et~al.}(2013){Burkhart}, {Lazarian}, {Ossenkopf}, \&
  {Stutzki}}]{burkhart2013b}
{Burkhart}, B., {Lazarian}, A., {Ossenkopf}, V., \& {Stutzki}, J. 2013, \apj,
  771, 123

\bibitem[{{Burkhart} {et~al.}(2010){Burkhart}, {Stanimirovi{\'c}}, {Lazarian},
  \& {Kowal}}]{burkhart2010}
{Burkhart}, B., {Stanimirovi{\'c}}, S., {Lazarian}, A., \& {Kowal}, G. 2010,
  \apj, 708, 1204

\bibitem[{{Cao} {et~al.}(2022){Cao}, {Qiu}, {Zhang}, \& {Li}}]{cao2022}
{Cao}, Y., {Qiu}, K., {Zhang}, Q., \& {Li}, G.-X. 2022, \apj, 927, 106

\bibitem[{{Cao} {et~al.}(2019){Cao}, {Qiu}, {Zhang}, {Wang}, {Hu}, \&
  {Liu}}]{cao2019}
{Cao}, Y., {Qiu}, K., {Zhang}, Q., {et~al.} 2019, \apjs, 241, 1

\bibitem[{{Cao} {et~al.}(2021){Cao}, {Qiu}, {Zhang}, {Wang}, \&
  {Xiao}}]{cao2021}
{Cao}, Y., {Qiu}, K., {Zhang}, Q., {Wang}, Y., \& {Xiao}, Y. 2021, \apjl, 918,
  L4

\bibitem[{{Collins} {et~al.}(2012){Collins}, {Kritsuk}, {Padoan}, {Li}, {Xu},
  {Ustyugov}, \& {Norman}}]{collins2012}
{Collins}, D.~C., {Kritsuk}, A.~G., {Padoan}, P., {et~al.} 2012, \apj, 750, 13

\bibitem[{{Downes} \& {Rinehart}(1966)}]{downes1966}
{Downes}, D. \& {Rinehart}, R. 1966, \apj, 144, 937

\bibitem[{{Draine}(2011)}]{draine2011}
{Draine}, B.~T. 2011, {Physics of the Interstellar and Intergalactic Medium}
  (Princeton Series in Astrophysics)

\bibitem[{{Elmegreen} \& {Scalo}(2004)}]{elmegreen2004}
{Elmegreen}, B.~G. \& {Scalo}, J. 2004, \araa, 42, 211

\bibitem[{{Falceta-Gon{\c{c}}alves} {et~al.}(2015){Falceta-Gon{\c{c}}alves},
  {Kowal}, {de Gouveia Dal Pino}, {Santos-Lima}, {Nakwacki}, \&
  {Lazarian}}]{falceta-goncalves2015}
{Falceta-Gon{\c{c}}alves}, D., {Kowal}, G., {de Gouveia Dal Pino}, E., {et~al.}
  2015, Highlights of Astronomy, 16, 406

\bibitem[{{Federrath} \& {Klessen}(2013)}]{federrath2013}
{Federrath}, C. \& {Klessen}, R.~S. 2013, \apj, 763, 51

\bibitem[{{Gieser} {et~al.}(2022){Gieser}, {Beuther}, {Semenov}, {Suri},
  {Soler}, {Linz}, {Syed}, {Henning}, {Feng}, {M{\"o}ller}, {Palau}, {Winters},
  {Beltr{\'a}n}, {Kuiper}, {Moscadelli}, {Klaassen}, {Urquhart}, {Peters},
  {Longmore}, {S{\'a}nchez-Monge}, {Galv{\'a}n-Madrid}, {Pudritz}, \&
  {Johnston}}]{gieser2022}
{Gieser}, C., {Beuther}, H., {Semenov}, D., {et~al.} 2022, \aap, 657, A3

\bibitem[{{Hacar} {et~al.}(2017){Hacar}, {Alves}, {Tafalla}, \&
  {Goicoechea}}]{hacar2017}
{Hacar}, A., {Alves}, J., {Tafalla}, M., \& {Goicoechea}, J.~R. 2017, \aap,
  602, L2

\bibitem[{{Hacar} {et~al.}(2020){Hacar}, {Bosman}, \& {van
  Dishoeck}}]{hacar2020}
{Hacar}, A., {Bosman}, A.~D., \& {van Dishoeck}, E.~F. 2020, \aap, 635, A4

\bibitem[{{Hacar} {et~al.}(2022){Hacar}, {Clark S.}, {Heitsch}, {Kainulainen},
  {Panopoulou G.}, {Seifried}, \& {R}}]{hacar2022}
{Hacar}, A., {Clark S.}, E., {Heitsch}, F., {et~al.} 2022, arXiv e-prints,
  arXiv:2203.09562

\bibitem[{{Hacar} {et~al.}(2013){Hacar}, {Tafalla}, {Kauffmann}, \&
  {Kov{\'a}cs}}]{hacar2013}
{Hacar}, A., {Tafalla}, M., {Kauffmann}, J., \& {Kov{\'a}cs}, A. 2013, \aap,
  554, A55

\bibitem[{{Hennebelle} \& {Falgarone}(2012)}]{hennebelle2012}
{Hennebelle}, P. \& {Falgarone}, E. 2012, \aapr, 20, 55

\bibitem[{{Hildebrand}(1983)}]{hildebrand1983}
{Hildebrand}, R.~H. 1983, \qjras, 24, 267

\bibitem[{{Jackson} {et~al.}(2006){Jackson}, {Rathborne}, {Shah}, {Simon},
  {Bania}, {Clemens}, {Chambers}, {Johnson}, {Dormody}, {Lavoie}, \&
  {Heyer}}]{jackson2006}
{Jackson}, J.~M., {Rathborne}, J.~M., {Shah}, R.~Y., {et~al.} 2006, \apjs, 163,
  145

\bibitem[{{Kauffmann} {et~al.}(2008){Kauffmann}, {Bertoldi}, {Bourke}, {Evans},
  \& {Lee}}]{kauffmann2008}
{Kauffmann}, J., {Bertoldi}, F., {Bourke}, T.~L., {Evans}, N.~J., I., \& {Lee},
  C.~W. 2008, \aap, 487, 993

\bibitem[{{Kirk} {et~al.}(2013){Kirk}, {Myers}, {Bourke}, {Gutermuth},
  {Hedden}, \& {Wilson}}]{kirk2013}
{Kirk}, H., {Myers}, P.~C., {Bourke}, T.~L., {et~al.} 2013, \apj, 766, 115

\bibitem[{{Koch} {et~al.}(2019){Koch}, {Rosolowsky}, {Boyden}, {Burkhart},
  {Ginsburg}, {Loeppky}, \& {Offner}}]{koch2019}
{Koch}, E.~W., {Rosolowsky}, E.~W., {Boyden}, R.~D., {et~al.} 2019, \aj, 158, 1

\bibitem[{{Kolmogorov}(1941)}]{kolmogorov1941}
{Kolmogorov}, A. 1941, Akademiia Nauk SSSR Doklady, 30, 301

\bibitem[{{K{\"o}rtgen} {et~al.}(2017){K{\"o}rtgen}, {Bovino}, {Schleicher},
  {Giannetti}, \& {Banerjee}}]{koertgen2017}
{K{\"o}rtgen}, B., {Bovino}, S., {Schleicher}, D. R.~G., {Giannetti}, A., \&
  {Banerjee}, R. 2017, \mnras, 469, 2602

\bibitem[{{Kumar} \& {Grave}(2007)}]{kumar2007}
{Kumar}, M.~S.~N. \& {Grave}, J.~M.~C. 2007, \aap, 472, 155

\bibitem[{{Kumar} {et~al.}(2020){Kumar}, {Palmeirim}, {Arzoumanian}, \&
  {Inutsuka}}]{kumar2020}
{Kumar}, M.~S.~N., {Palmeirim}, P., {Arzoumanian}, D., \& {Inutsuka}, S.~I.
  2020, \aap, 642, A87

\bibitem[{{Larson}(2003)}]{larson2003}
{Larson}, R.~B. 2003, Reports on Progress in Physics, 66, 1651

\bibitem[{{Lazarian} \& {Pogosyan}(2000)}]{lazarian2000}
{Lazarian}, A. \& {Pogosyan}, D. 2000, \apj, 537, 720

\bibitem[{{Lazarian} \& {Pogosyan}(2004)}]{lazarian2004}
{Lazarian}, A. \& {Pogosyan}, D. 2004, \apj, 616, 943

\bibitem[{{Lazarian} {et~al.}(2002){Lazarian}, {Pogosyan}, \&
  {Esquivel}}]{lazarian2002}
{Lazarian}, A., {Pogosyan}, D., \& {Esquivel}, A. 2002, in Astronomical Society
  of the Pacific Conference Series, Vol. 276, Seeing Through the Dust: The
  Detection of HI and the Exploration of the ISM in Galaxies, ed. A.~R.
  {Taylor}, T.~L. {Landecker}, \& A.~G. {Willis}, 182

\bibitem[{{Li} {et~al.}(2021){Li}, {Cao}, \& {Qiu}}]{li2021}
{Li}, G.-X., {Cao}, Y., \& {Qiu}, K. 2021, \apj, 916, 13

\bibitem[{{Marsh} {et~al.}(2015){Marsh}, {Whitworth}, \& {Lomax}}]{marsh2015}
{Marsh}, K.~A., {Whitworth}, A.~P., \& {Lomax}, O. 2015, \mnras, 454, 4282

\bibitem[{{Marsh} {et~al.}(2017){Marsh}, {Whitworth}, {Lomax}, {Ragan},
  {Becciani}, {Cambr{\'e}sy}, {Di Giorgio}, {Eden}, {Elia}, {Kacsuk},
  {Molinari}, {Palmeirim}, {Pezzuto}, {Schneider}, {Sciacca}, \&
  {Vitello}}]{marsh2017}
{Marsh}, K.~A., {Whitworth}, A.~P., {Lomax}, O., {et~al.} 2017, \mnras, 471,
  2730

\bibitem[{{Minier} {et~al.}(2003){Minier}, {Ellingsen}, {Norris}, \&
  {Booth}}]{minier2003}
{Minier}, V., {Ellingsen}, S.~P., {Norris}, R.~P., \& {Booth}, R.~S. 2003,
  \aap, 403, 1095

\bibitem[{{Miville-Desch{\^e}nes} {et~al.}(2003){Miville-Desch{\^e}nes},
  {Joncas}, {Falgarone}, \& {Boulanger}}]{miville-deschenes2003}
{Miville-Desch{\^e}nes}, M.~A., {Joncas}, G., {Falgarone}, E., \& {Boulanger},
  F. 2003, \aap, 411, 109

\bibitem[{{Molinari} {et~al.}(2016{\natexlab{a}}){Molinari}, {Merello}, {Elia},
  {Cesaroni}, {Testi}, \& {Robitaille}}]{molinari2016}
{Molinari}, S., {Merello}, M., {Elia}, D., {et~al.} 2016{\natexlab{a}}, \apjl,
  826, L8

\bibitem[{{Molinari} {et~al.}(2016{\natexlab{b}}){Molinari}, {Schisano},
  {Elia}, {Pestalozzi}, {Traficante}, {Pezzuto}, {Swinyard}, {Noriega-Crespo},
  {Bally}, {Moore}, {Plume}, {Zavagno}, {di Giorgio}, {Liu}, {Pilbratt},
  {Mottram}, {Russeil}, {Piazzo}, {Veneziani}, {Benedettini}, {Calzoletti},
  {Faustini}, {Natoli}, {Piacentini}, {Merello}, {Palmese}, {Del Grande},
  {Polychroni}, {Rygl}, {Polenta}, {Barlow}, {Bernard}, {Martin}, {Testi},
  {Ali}, {Andr{\'e}}, {Beltr{\'a}n}, {Billot}, {Carey}, {Cesaroni},
  {Compi{\`e}gne}, {Eden}, {Fukui}, {Garcia-Lario}, {Hoare}, {Huang}, {Joncas},
  {Lim}, {Lord}, {Martinavarro-Armengol}, {Motte}, {Paladini}, {Paradis},
  {Peretto}, {Robitaille}, {Schilke}, {Schneider}, {Schulz}, {Sibthorpe},
  {Strafella}, {Thompson}, {Umana}, {Ward-Thompson}, \&
  {Wyrowski}}]{molinari2016b}
{Molinari}, S., {Schisano}, E., {Elia}, D., {et~al.} 2016{\natexlab{b}}, \aap,
  591, A149

\bibitem[{{Molinari} {et~al.}(2010){Molinari}, {Swinyard}, {Bally}, {Barlow},
  {Bernard}, {Martin}, {Moore}, {Noriega-Crespo}, {Plume}, {Testi}, {Zavagno},
  {Abergel}, {Ali}, {Anderson}, {Andr{\'e}}, {Baluteau}, {Battersby},
  {Beltr{\'a}n}, {Benedettini}, {Billot}, {Blommaert}, {Bontemps}, {Boulanger},
  {Brand}, {Brunt}, {Burton}, {Calzoletti}, {Carey}, {Caselli}, {Cesaroni},
  {Cernicharo}, {Chakrabarti}, {Chrysostomou}, {Cohen}, {Compiegne}, {de
  Bernardis}, {de Gasperis}, {di Giorgio}, {Elia}, {Faustini}, {Flagey},
  {Fukui}, {Fuller}, {Ganga}, {Garcia-Lario}, {Glenn}, {Goldsmith}, {Griffin},
  {Hoare}, {Huang}, {Ikhenaode}, {Joblin}, {Joncas}, {Juvela}, {Kirk},
  {Lagache}, {Li}, {Lim}, {Lord}, {Marengo}, {Marshall}, {Masi}, {Massi},
  {Matsuura}, {Minier}, {Miville-Desch{\^e}nes}, {Montier}, {Morgan}, {Motte},
  {Mottram}, {M{\"u}ller}, {Natoli}, {Neves}, {Olmi}, {Paladini}, {Paradis},
  {Parsons}, {Peretto}, {Pestalozzi}, {Pezzuto}, {Piacentini}, {Piazzo},
  {Polychroni}, {Pomar{\`e}s}, {Popescu}, {Reach}, {Ristorcelli}, {Robitaille},
  {Robitaille}, {Rod{\'o}n}, {Roy}, {Royer}, {Russeil}, {Saraceno}, {Sauvage},
  {Schilke}, {Schisano}, {Schneider}, {Schuller}, {Schulz}, {Sibthorpe},
  {Smith}, {Smith}, {Spinoglio}, {Stamatellos}, {Strafella}, {Stringfellow},
  {Sturm}, {Taylor}, {Thompson}, {Traficante}, {Tuffs}, {Umana}, {Valenziano},
  {Vavrek}, {Veneziani}, {Viti}, {Waelkens}, {Ward-Thompson}, {White},
  {Wilcock}, {Wyrowski}, {Yorke}, \& {Zhang}}]{molinari2010}
{Molinari}, S., {Swinyard}, B., {Bally}, J., {et~al.} 2010, \aap, 518, L100

\bibitem[{{Motte} {et~al.}(2018){Motte}, {Bontemps}, \& {Louvet}}]{motte2018}
{Motte}, F., {Bontemps}, S., \& {Louvet}, F. 2018, \araa, 56, 41

\bibitem[{{Motte} {et~al.}(2007){Motte}, {Bontemps}, {Schilke}, {Schneider},
  {Menten}, \& {Brogui{\`e}re}}]{motte2007}
{Motte}, F., {Bontemps}, S., {Schilke}, P., {et~al.} 2007, \aap, 476, 1243

\bibitem[{{Motte} {et~al.}(2010){Motte}, {Zavagno}, {Bontemps}, {Schneider},
  {Hennemann}, {di Francesco}, {Andr{\'e}}, {Saraceno}, {Griffin}, {Marston},
  {Ward-Thompson}, {White}, {Minier}, {Men'shchikov}, {Hill}, {Abergel},
  {Anderson}, {Aussel}, {Balog}, {Baluteau}, {Bernard}, {Cox}, {Csengeri},
  {Deharveng}, {Didelon}, {di Giorgio}, {Hargrave}, {Huang}, {Kirk}, {Leeks},
  {Li}, {Martin}, {Molinari}, {Nguyen-Luong}, {Olofsson}, {Persi}, {Peretto},
  {Pezzuto}, {Roussel}, {Russeil}, {Sadavoy}, {Sauvage}, {Sibthorpe},
  {Spinoglio}, {Testi}, {Teyssier}, {Vavrek}, {Wilson}, \&
  {Woodcraft}}]{motte2010}
{Motte}, F., {Zavagno}, A., {Bontemps}, S., {et~al.} 2010, \aap, 518, L77+

\bibitem[{{Myers} {et~al.}(1996){Myers}, {Mardones}, {Tafalla}, {Williams}, \&
  {Wilner}}]{myers1996}
{Myers}, P.~C., {Mardones}, D., {Tafalla}, M., {Williams}, J.~P., \& {Wilner},
  D.~J. 1996, \apjl, 465, L133

\bibitem[{{Oliveira} {et~al.}(2003){Oliveira}, {H{\'e}brard}, {Howk}, {Kruk},
  {Chayer}, \& {Moos}}]{oliveira2003}
{Oliveira}, C.~M., {H{\'e}brard}, G., {Howk}, J.~C., {et~al.} 2003, \apj, 587,
  235

\bibitem[{{Ortiz-Le{\'o}n} {et~al.}(2021){Ortiz-Le{\'o}n}, {Menten},
  {Brunthaler}, {Csengeri}, {Urquhart}, {Wyrowski}, {Gong}, {Rugel}, {Dzib},
  {Yang}, {Nguyen}, {Cotton}, {Medina}, {Dokara}, {K{\"o}nig}, {Beuther},
  {Pandian}, {Reich}, \& {Roy}}]{ortiz-leon2021}
{Ortiz-Le{\'o}n}, G.~N., {Menten}, K.~M., {Brunthaler}, A., {et~al.} 2021,
  \aap, 651, A87

\bibitem[{{Ossenkopf} \& {Henning}(1994)}]{ossenkopf1994}
{Ossenkopf}, V. \& {Henning}, T. 1994, \aap, 291, 943

\bibitem[{{Panopoulou} {et~al.}(2017){Panopoulou}, {Psaradaki}, {Skalidis},
  {Tassis}, \& {Andrews}}]{panopoulou2017}
{Panopoulou}, G.~V., {Psaradaki}, I., {Skalidis}, R., {Tassis}, K., \&
  {Andrews}, J.~J. 2017, \mnras, 466, 2529

\bibitem[{{Parise} {et~al.}(2009){Parise}, {Leurini}, {Schilke}, {Roueff},
  {Thorwirth}, \& {Lis}}]{parise2009}
{Parise}, B., {Leurini}, S., {Schilke}, P., {et~al.} 2009, \aap, 508, 737

\bibitem[{{Peretto} {et~al.}(2014){Peretto}, {Fuller}, {Andr{\'e}},
  {Arzoumanian}, {Rivilla}, {Bardeau}, {Duarte Puertas}, {Guzman Fernandez},
  {Lenfestey}, {Li}, {Olguin}, {R{\"o}ck}, {de Villiers}, \&
  {Williams}}]{peretto2014}
{Peretto}, N., {Fuller}, G.~A., {Andr{\'e}}, P., {et~al.} 2014, \aap, 561, A83

\bibitem[{{Pety} {et~al.}(2007){Pety}, {Goicoechea}, {Hily-Blant}, {Gerin}, \&
  {Teyssier}}]{pety2007}
{Pety}, J., {Goicoechea}, J.~R., {Hily-Blant}, P., {Gerin}, M., \& {Teyssier},
  D. 2007, \aap, 464, L41

\bibitem[{{Pillai} {et~al.}(2012){Pillai}, {Caselli}, {Kauffmann}, {Zhang},
  {Thompson}, \& {Lis}}]{pillai2012}
{Pillai}, T., {Caselli}, P., {Kauffmann}, J., {et~al.} 2012, \apj, 751, 135

\bibitem[{{Pingel} {et~al.}(2018){Pingel}, {Lee}, {Burkhart}, \&
  {Stanimirovi{\'c}}}]{pingel2018}
{Pingel}, N.~M., {Lee}, M.-Y., {Burkhart}, B., \& {Stanimirovi{\'c}}, S. 2018,
  \apj, 856, 136

\bibitem[{{Reipurth}(2008)}]{reipurth2008}
{Reipurth}, B., ed. 2008, {Star Formation and Young Clusters in Cygnus}, ed.
  B.~{Reipurth}, Vol.~4, 36

\bibitem[{{Rygl} {et~al.}(2012){Rygl}, {Brunthaler}, {Sanna}, {Menten}, {Reid},
  {van Langevelde}, {Honma}, {Torstensson}, \& {Fujisawa}}]{rygl2012}
{Rygl}, K.~L.~J., {Brunthaler}, A., {Sanna}, A., {et~al.} 2012, \aap, 539, A79

\bibitem[{{Schneider} {et~al.}(2006){Schneider}, {Bontemps}, {Simon}, {Jakob},
  {Motte}, {Miller}, {Kramer}, \& {Stutzki}}]{schneider2006}
{Schneider}, N., {Bontemps}, S., {Simon}, R., {et~al.} 2006, \aap, 458, 855

\bibitem[{{Schneider} {et~al.}(2010){Schneider}, {Csengeri}, {Bontemps},
  {Motte}, {Simon}, {Hennebelle}, {Federrath}, \& {Klessen}}]{schneider2010}
{Schneider}, N., {Csengeri}, T., {Bontemps}, S., {et~al.} 2010, \aap, 520, A49

\bibitem[{{Schuller} {et~al.}(2009){Schuller}, {Menten}, {Contreras},
  {Wyrowski}, {Schilke}, {Bronfman}, {Henning}, {Walmsley}, {Beuther},
  {Bontemps}, {Cesaroni}, {Deharveng}, {Garay}, {Herpin}, {Lefloch}, {Linz},
  {Mardones}, {Minier}, {Molinari}, {Motte}, {Nyman}, {Reveret}, {Risacher},
  {Russeil}, {Schneider}, {Testi}, {Troost}, {Vasyunina}, {Wienen}, {Zavagno},
  {Kovacs}, {Kreysa}, {Siringo}, \& {Wei{\ss}}}]{schuller2009}
{Schuller}, F., {Menten}, K.~M., {Contreras}, Y., {et~al.} 2009, \aap, 504, 415

\bibitem[{{Shirley}(2015)}]{shirley2015}
{Shirley}, Y.~L. 2015, \pasp, 127, 299

\bibitem[{{Sousbie}(2011)}]{sousbie2011a}
{Sousbie}, T. 2011, \mnras, 414, 350

\bibitem[{{Sousbie} {et~al.}(2011){Sousbie}, {Pichon}, \&
  {Kawahara}}]{sousbie2011b}
{Sousbie}, T., {Pichon}, C., \& {Kawahara}, H. 2011, \mnras, 414, 384

\bibitem[{{Stahler} \& {Palla}(2005)}]{stahler2005}
{Stahler}, S.~W. \& {Palla}, F. 2005, {The Formation of Stars} (ISBN
  3-527-40559-3.~Wiley-VCH)

\bibitem[{{Stanimirovi{\'c}} \& {Lazarian}(2001)}]{stanimirovic2001}
{Stanimirovi{\'c}}, S. \& {Lazarian}, A. 2001, \apjl, 551, L53

\bibitem[{{Suri} {et~al.}(2019){Suri}, {S{\'a}nchez-Monge}, {Schilke},
  {Clarke}, {Smith}, {Ossenkopf-Okada}, {Klessen}, {Padoan}, {Goldsmith},
  {Arce}, {Bally}, {Carpenter}, {Ginsburg}, {Johnstone}, {Kauffmann}, {Kong},
  {Lis}, {Mairs}, {Pillai}, {Pineda}, \& {Duarte-Cabral}}]{suri2019}
{Suri}, S., {S{\'a}nchez-Monge}, {\'A}., {Schilke}, P., {et~al.} 2019, \aap,
  623, A142

\bibitem[{{Taylor} {et~al.}(2003){Taylor}, {Gibson}, {Peracaula}, {Martin},
  {Landecker}, {Brunt}, {Dewdney}, {Dougherty}, {Gray}, {Higgs}, {Kerton},
  {Knee}, {Kothes}, {Purton}, {Uyaniker}, {Wallace}, {Willis}, \&
  {Durand}}]{taylor2003}
{Taylor}, A.~R., {Gibson}, S.~J., {Peracaula}, M., {et~al.} 2003, \aj, 125,
  3145

\bibitem[{{Tobin} {et~al.}(2012){Tobin}, {Hartmann}, {Bergin}, {Chiang},
  {Looney}, {Chandler}, {Maret}, \& {Heitsch}}]{tobin2012}
{Tobin}, J.~J., {Hartmann}, L., {Bergin}, E., {et~al.} 2012, \apj, 748, 16

\bibitem[{{van der Walt} {et~al.}(2021){van der Walt}, {Kristensen},
  {J{\o}rgensen}, {Calcutt}, {Manigand}, {el Akel}, {Garrod}, \&
  {Qiu}}]{vanderwalt2021}
{van der Walt}, S.~J., {Kristensen}, L.~E., {J{\o}rgensen}, J.~K., {et~al.}
  2021, \aap, 655, A86

\bibitem[{{V{\'a}zquez-Semadeni} {et~al.}(2009){V{\'a}zquez-Semadeni},
  {G{\'o}mez}, {Jappsen}, {Ballesteros-Paredes}, \& {Klessen}}]{vazquez2009}
{V{\'a}zquez-Semadeni}, E., {G{\'o}mez}, G.~C., {Jappsen}, A.~K.,
  {Ballesteros-Paredes}, J., \& {Klessen}, R.~S. 2009, \apj, 707, 1023

\bibitem[{{V{\'a}zquez-Semadeni} {et~al.}(2019){V{\'a}zquez-Semadeni}, {Palau},
  {Ballesteros-Paredes}, {G{\'o}mez}, \& {Zamora-Avil{\'e}s}}]{vazquez2019}
{V{\'a}zquez-Semadeni}, E., {Palau}, A., {Ballesteros-Paredes}, J.,
  {G{\'o}mez}, G.~C., \& {Zamora-Avil{\'e}s}, M. 2019, \mnras, 490, 3061

\bibitem[{{von Weizs{\"a}cker}(1951)}]{weizsaecker1951}
{von Weizs{\"a}cker}, C.~F. 1951, \apj, 114, 165

\bibitem[{{Walsh} {et~al.}(2006){Walsh}, {Bourke}, \& {Myers}}]{walsh2006}
{Walsh}, A.~J., {Bourke}, T.~L., \& {Myers}, P.~C. 2006, \apj, 637, 860

\bibitem[{{Wang} {et~al.}(2020){Wang}, {Beuther}, {Rugel}, {Soler}, {Stil},
  {Ott}, {Bihr}, {McClure-Griffiths}, {Anderson}, {Klessen}, {Goldsmith},
  {Roy}, {Glover}, {Urquhart}, {Heyer}, {Linz}, {Smith}, {Bigiel}, {Dempsey},
  \& {Henning}}]{wang2020a}
{Wang}, Y., {Beuther}, H., {Rugel}, M.~R., {et~al.} 2020, \aap, 634, A83

\bibitem[{{Zhou}(1992)}]{zhou1992}
{Zhou}, S. 1992, \apj, 394, 204

\end{thebibliography}

\newpage
\begin{appendix}

\section{Additional figures and line table}

\begin{figure*}[h]
\includegraphics[width=0.9\textwidth]{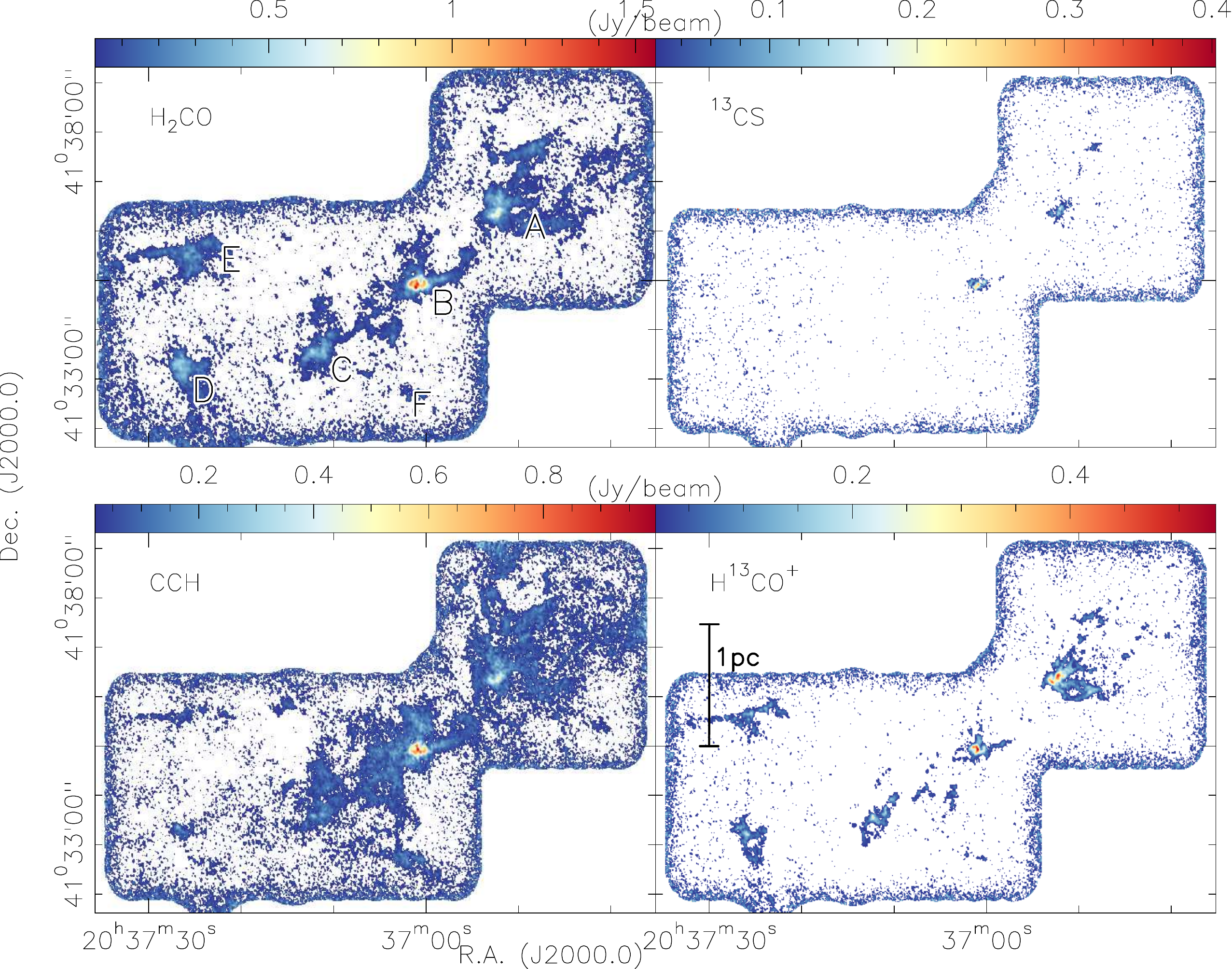}
\includegraphics[width=0.9\textwidth]{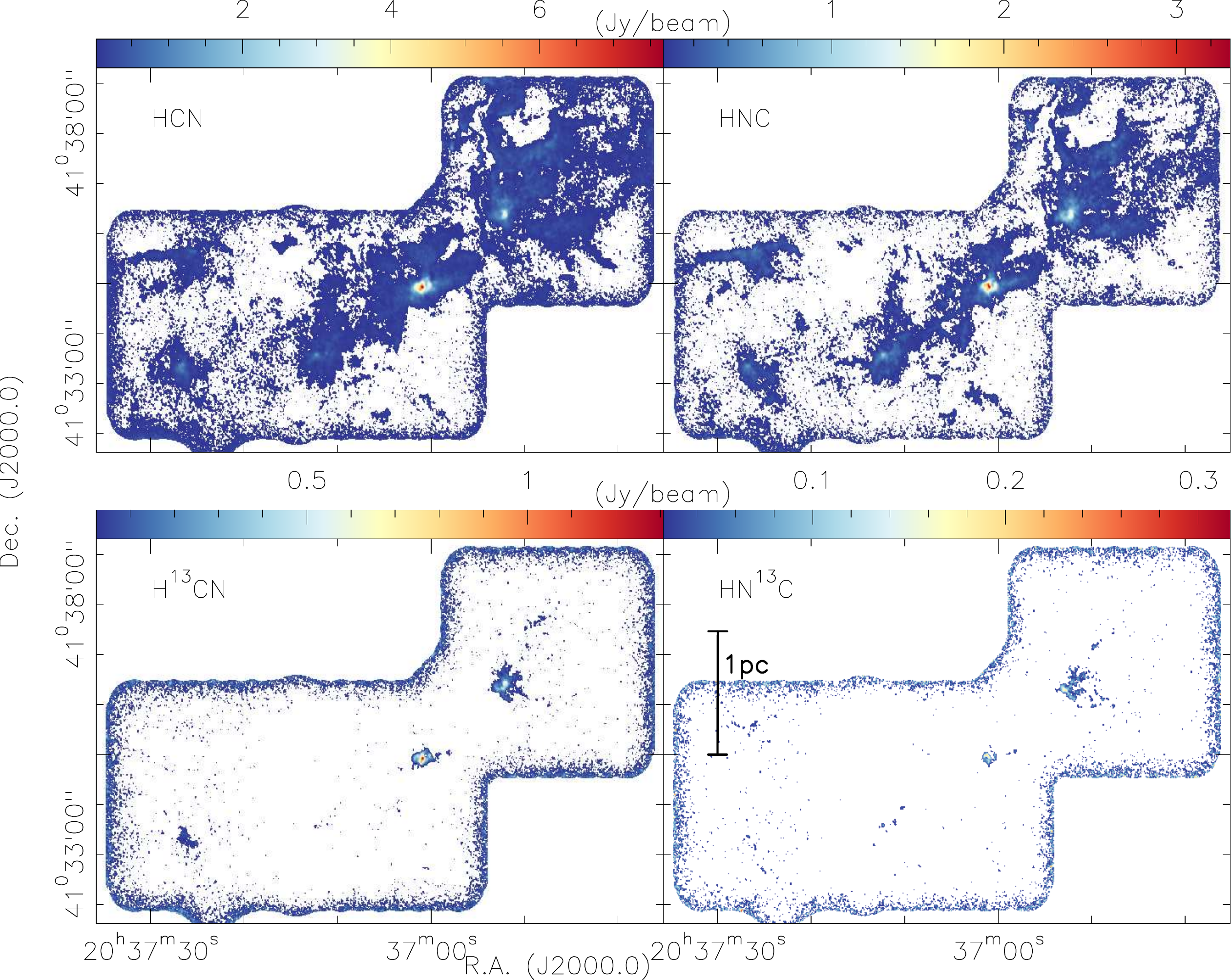}
\caption{Integrated intensities for the species marked in each
  panel. The integration ranges are always from $-7$ to
  3\,km\,s$^{-1}$. Only for HCN and H$^{13}$CN were broader integration
  ranges from $-14$ to 8\,km\,s$^{-1}$ used to cover all
  hyperfine structures of the line. The data were clipped below the
  $4\sigma$ level for each species (Table \ref{rms}).}
\label{mom0_2} 
\end{figure*} 

\begin{figure*}[htb]
\includegraphics[width=0.90\textwidth]{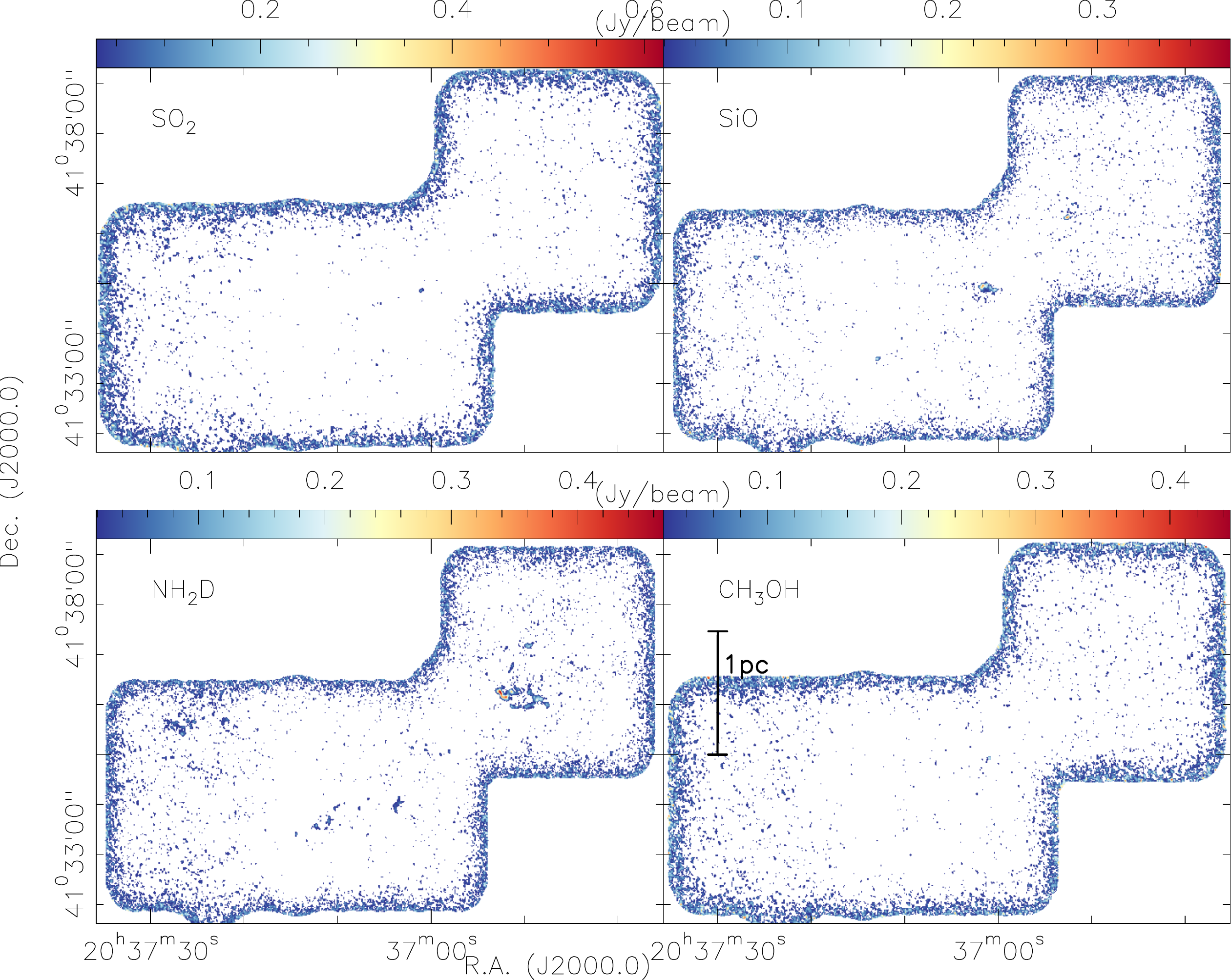}
\includegraphics[width=0.90\textwidth]{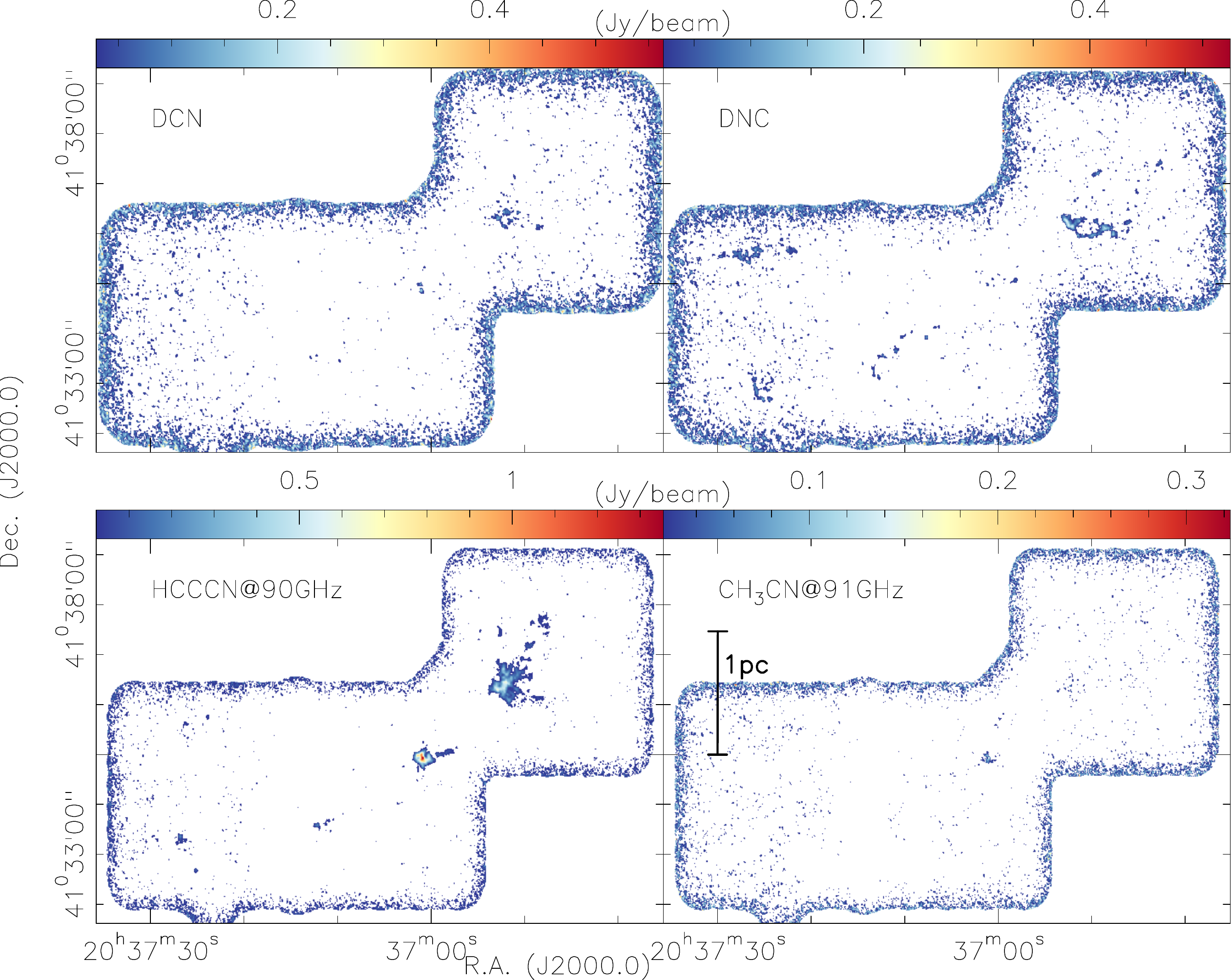}
\caption{Integrated intensities for the species marked in each
  panel. The integration ranges are always from -7 to
  3\,km\,s$^{-1}$. The data were clipped below the
  $4\sigma$ level for each species (Table \ref{rms}).}
\label{mom0_3} 
\end{figure*} 

\begin{figure*}[htb]
\includegraphics[width=0.99\textwidth]{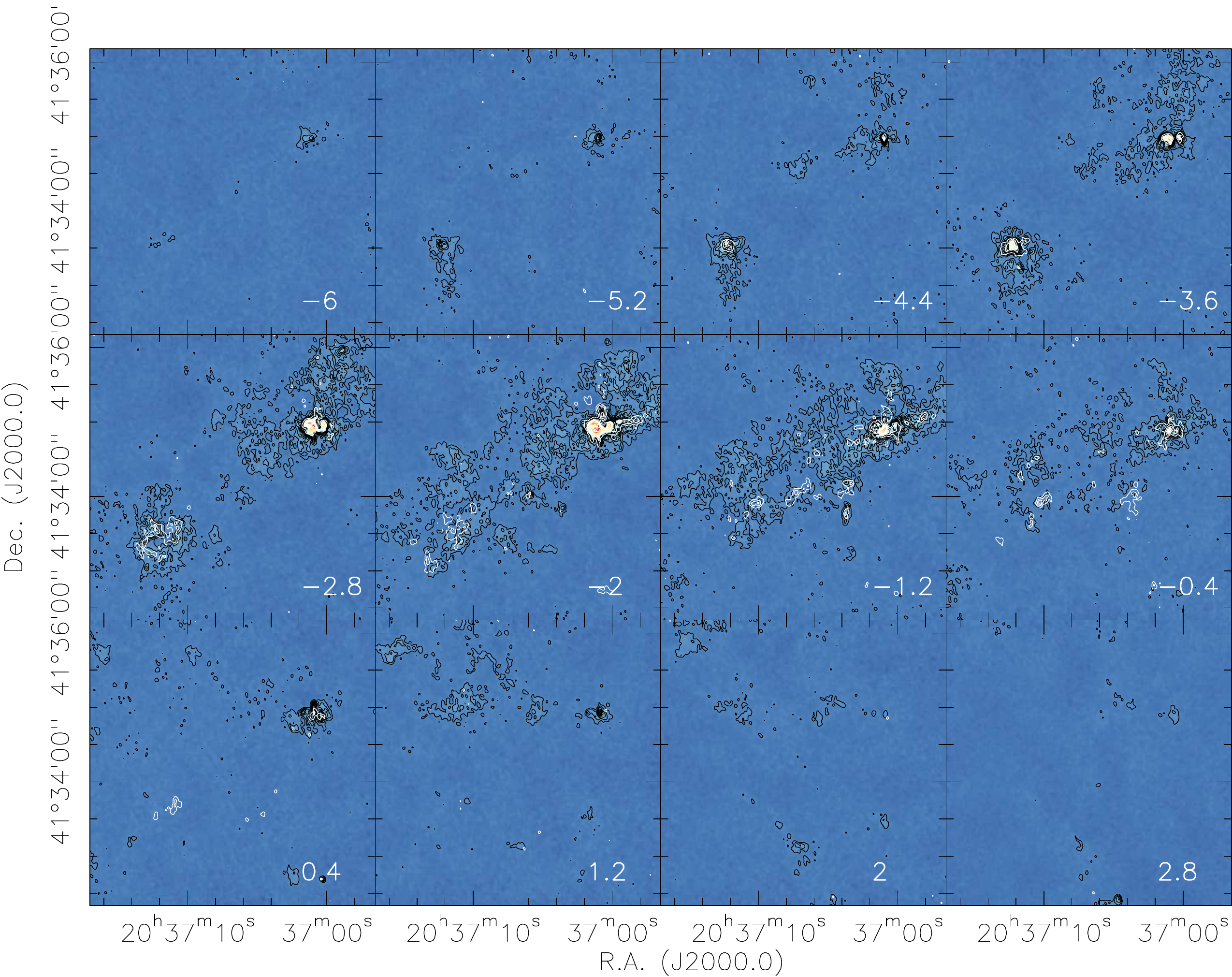}
\caption{Channel maps of the central third of DR20 in the HCO$^+(1-0)$ and
  H$^{13}$CO$^+(1-0)$ emission lines. The color and black contours
  show the HCO$^+(1-0)$ emission, whereas the white contours present
  the H$^{13}$CO$^+(1-0)$. Contour levels are always in 5$\sigma$
  steps, and the central velocity is marked in each panel.}
\label{hco+_channel2} 
\end{figure*}

\begin{table*}[htb]
\caption{Spectral correlator settings with targeted lines.}
\begin{tabular}{lrrrl}
  \hline \hline
cent.~freq. & SB & width & $\Delta \varv$ & frequencies and molecules \\
(GHz) &    & (MHz) & (MHz) & (MHz) \\
\hline
72.246 & LSB & 4064 & 2.00 & wideband \\
76.307 & LSB & 4064 & 2.00 & wideband \\
87.733 & USB & 4064 & 2.00 & wideband \\
91.795 & USB & 4064 & 2.00 & wideband \\
71.909 & LSB & 64   & 0.06 & 71889.741 HC$_5$N \\
72.037 & LSB & 64   & 0.06 & 72039.303 DCO$^+$ \\
72.101 & LSB & 64   & 0.06 & 72107.7205 CCD \\
72.165 & LSB & 64   & 0.06 & 72156.5915527 72177.6297913 \\
72.421 & LSB & 64   & 0.06 & 72414.905 DCN, 72409.090 H$_2$CO \\
72.485 & LSB & 64   & 0.06 & 72475.075 HC$^{13}$CCN, 72482.060 HCC$^{13}$CN, 72409.090 H$_2$CO \\
72.741 & LSB & 64   & 0.06 & 72758.243 SO$_2$ \\
72.805 & LSB & 64   & 0.06 & 72783.822 HCCCN, 72837.948 H$_2$CO \\
72.869 & LSB & 64   & 0.06 & 72837.948 H$_2$CO \\
72.996 & LSB & 64   & 0.06 & 72976.779 OCS \\
73.572 & LSB & 64   & 0.06 & 73590.218 CH$_3$CN \\
74.084 & LSB & 64   & 0.06 & 74111.42 HCNH$^+$ \\
74.148 & LSB & 64   & 0.06 & 74111.42 HCNH$^+$ \\
74.660 & LSB & 64   & 0.06 & 74644.568 H44$\alpha$ \\
74.852 & LSB & 64   & 0.06 & 74869.0231934 HC5N \\
74.916 & LSB & 64   & 0.06 & 74891.677 74924.134 CH$_3$CHO, 74939.624 H55$\beta$ \\
75.812 & LSB & 64   & 0.06 & 75816.443 CH$_3$SH \\
75.876 & LSB & 64   & 0.06 & 75864.405 75862.859 CH$_3$SH \\
75.940 & LSB & 64   & 0.06 & 75925.914 CH$_3$SH \\
76.132 & LSB & 64   & 0.06 & 76117.432 76156.028 C$_4$H \\
76.324 & LSB & 64   & 0.06 & 76305.727 DNC \\
76.516 & LSB & 64   & 0.06 & 76509.684 CH$_3$OH \\
76.836 & LSB & 64   & 0.06 & 76838.932 CH$_3$NH$_2$ \\
76.900 & LSB & 64   & 0.06 & 76878.952 CH$_3$CHO \\
77.028 & LSB & 64   & 0.06 & 77038.605 CH$_3$CHO \\
77.092 & LSB & 64   & 0.06 & 77082.951 CH$_3$CHO, 77109.632 N$_2$D$^+$ \\
77.156 & LSB & 64   & 0.06 & 77125.695 CH$_3$CHO \\
77.220 & LSB & 64   & 0.06 & 77214.359 HC$_5$N \\
77.732 & LSB & 64   & 0.06 & 77731.711 CCS, \\
85.924 & USB & 64   & 0.06 & 85926.270 NH$_2$D \\
86.052 & USB & 64   & 0.06 & 86054.961 HC$^{15}$N \\
86.308 & USB & 64   & 0.06 & 86338.7367 86340.1764 86342.2551 H$^{13}$CN \\ 
86.372 & USB & 64   & 0.06 & 86338.7367 86340.1764 86342.2551 H$^{13}$CN \\
86.692 & USB & 64   & 0.06 & 86670.760 86708.360 HCO \\
86.756 & USB & 64   & 0.06 & 86754.288 H$^{13}$CO$^+$ \\
86.820 & USB & 64   & 0.06 & 86846.960 SiO \\
86.884 & USB & 64   & 0.06 & 87863.916 HC$_5$N \\
87.076 & USB & 64   & 0.06 & 87090.735 87090.859 87090.942 HN$^{13}$C \\
87.332 & USB & 64   & 0.06 & 87316.925 87328.624 CCH \\
87.396 & USB & 64   & 0.06 & 87402.004 87407.165 CCH \\
87.844 & USB & 64   & 0.06 & 87848.855 NH$_2$CHO \\
87.908 & USB & 64   & 0.06 & 87925.252 HNCO \\
88.612 & USB & 64   & 0.06 & 88631.847 HCN \\
88.868 & USB & 64   & 0.06 & 88851.607 88843.187 CH$_3$OCHO, 88865.715 H$^{15}$NC \\
89.188 & USB & 64   & 0.06 & 89188.52470 HCO$^+$, 88166.806 H$^{13}$CCCN \\
90.532 & USB & 64   & 0.06 & 90526.205 HC$_5$N \\
90.596 & USB & 64   & 0.06 & 90593.061 HC$^{13}$CCN \\
90.660 & USB & 64   & 0.06 & 90663.593 HNC, 90686.381 CCS \\
90.788 & USB & 64   & 0.06 & 90771.550 SiS \\
90.980 & USB & 64   & 0.06 & 90979.023 HCCCN \\
91.172 & USB & 64   & 0.06 & 91171.039 HDCS \\
91.940 & USB & 64   & 0.06 & 91987.088 CH$_3$CN \\
92.004 & USB & 64   & 0.06 & 91987.088 CH$_3$CN \\
92.068 & USB & 64   & 0.06 & 92070.28 HCCCCN, 92075.51 CH$_3$OD \\ 
92.452 & USB & 64   & 0.06 & 92426.248 CH$_2$CHCN, 92494.308 $^{13}$CS \\
92.516 & USB & 64   & 0.06 & 92494.308 $^{13}$CS \\
93.156 & USB & 64   & 0.06 & 93173.777 N$_2$H$^+$ \\
\hline \hline
\end{tabular}
\label{lines}
\end{table*}

\end{appendix}

\end{document}